\documentclass[preprint]{ptephy_v1}

\preprintnumber{XXXX-XXXX}

\usepackage{amsmath}
\usepackage{hyperref}
\usepackage{graphicx}
\usepackage{subcaption}
\usepackage{array}
\usepackage{booktabs}
\usepackage{tensor}
\usepackage{multirow}
\usepackage{url}
\usepackage[utf8]{inputenc}
\usepackage[usenames,dvipsnames]{xcolor}
\usepackage{ulem}

\newcolumntype{P}[1]{>{\centering\arraybackslash}p{#1}}
\newcolumntype{M}[1]{>{\centering\arraybackslash}m{#1}}

\begin{document}


\title{A superconducting tensor detector for mid-frequency gravitational waves: its multi-channel nature and main astrophysical targets}

\author{Yeong-Bok Bae\thanks{These authors contributed equally to this work}}
\affil{Particle Theory and Cosmology Group, Center for Theoretical Physics of the Universe, Institute for Basic Science (IBS), Daejeon 34126, Korea}

\author[1]{Chan Park\textsuperscript{\dag}}

\author{Edwin J. Son\textsuperscript{\dag}}
\affil{National Institute for Mathematical Sciences, Daejeon 34047, Korea}

\author{Sang-Hyeon Ahn}
\affil{Korea Astronomy and Space Science Institute, Daejeon 34055, Korea}

\author{Minjoong Jeong}
\affil{Supercomputing Center, Korea Institute of Science and Technology Information, Daejeon 34141, Korea}

\author{Gungwon Kang\thanks{Corresponding authors: gwkang@cau.ac.kr, chunglee.kim@ewha.ac.kr}}
\affil{Chung-Ang University, Seoul 06974, Korea \email{gwkang@cau.ac.kr}}

\author{Chunglee Kim\textsuperscript{\ddag}}
\affil{Department of Physics, Ewha Womans University, Seoul 03760, Korea \email{chunglee.kim@ewha.ac.kr}}

\author{Dong Lak Kim}
\affil{Korea Basic Science Institute, Daejeon 34051, Korea}

\author{Jaewan Kim}
\affil{Department of Physics, Myongji University, Yongin 17058, Korea}

\author[2]{Whansun Kim}

\author{Hyung Mok Lee}
\affil{Astronomy Program Department of Physics and Astronomy, Seoul National University, Seoul 08826, Korea}

\author{Yong-Ho Lee}
\affil{Korea Research Institute of Standards and Science, Daejeon 305-600, Korea}

\author{Ronald S. Norton}
\affil{Department of Physics, University of Maryland, College Park, MD 20742, USA}

\author[2]{John J. Oh}

\author[2]{Sang Hoon Oh}

\author[11]{Ho Jung Paik}


\begin{abstract}

Mid-frequency band gravitational-wave detectors will be complementary for the existing Earth-based detectors (sensitive above 10 Hz or so) and the future space-based detectors such as LISA, which will be sensitive below around 10 mHz. A ground-based superconducting omnidirectional gravitational radiation observatory (SOGRO) has recently been proposed along with several design variations for the frequency band of 0.1 to 10 Hz. For three conceptual designs of SOGRO ({\it e.g.,} pSOGRO, SOGRO and aSOGRO), we examine their multi-channel natures, sensitivities and science cases. One of the key characteristics of the SOGRO concept is its six detection channels. The response functions of each channel are calculated for all possible gravitational wave polarizations including scalar and vector modes. Combining these response functions, we also confirm the omnidirectional nature of SOGRO. Hence, even a single SOGRO detector will be able to determine the position of a source and polarizations of gravitational waves, if detected. Taking into account SOGRO's sensitivity and technical requirements, two main targets are most plausible: gravitational waves from compact binaries and stochastic backgrounds. Based on assumptions we consider in this work, detection rates for intermediate-mass binary black holes (in the mass range of hundreds up to $10^{4}$ $M_\odot$) are expected to be $0.0014-2.5 \,\, {\rm yr}^{-1}$. In order to detect stochastic gravitational wave background, multiple detectors are required. Two aSOGRO detector networks may be able to put limits on the stochastic background beyond the indirect limit from cosmological observations.

\end{abstract}

\subjectindex{Gravitational waves, Observational astronomy, Black holes, Superconducting, Cryogenic}

\maketitle


\section{Introduction}

\noindent
Since the detection of GW150914 \cite{Abbott:GW150914}, the gravitational wave (GW) frequency band between $20-2000$ Hz has been shown to be fruitful in searching for coalescence of compact binaries composed of stellar-mass black holes (BHs) or neutron stars (NSs).
Observations of about 90 such GW events up to the third observing run have been done, opening up a totally new way of exploring the universe.
All ground-based laser interferometers such as the advanced Laser Interferometer Gravitational-wave Observatory (aLIGO) \cite{aLIGO:2015}, advanced Virgo \cite{aVirgo:2015} and KAGRA \cite{KAGRA:2019} are optimized in this frequency bandwidth.

To broaden the frequency bandwidth, several efforts are underway. These include the Pulsar Timing Arrays (PTAs) \cite{EPTA3:2023,EPTA4:2023,PPTA:2023,NANOGrav:2023,CPTA:2023} in the nanohertz GW frequency band, the Laser Interferometer Space Antenna (LISA) in $10^{-5}-10^{-1}$ Hz range \cite{LISA:2017}, and the Einstein Telescope (ET) \cite{ET:2010} underground that reaches down to 1 Hz. Different frequency bands are targeted to search for different types of GW sources.

Within the past decade, the mid-frequency band between $0.1-10$ Hz has increasingly drawn attention in the GW community~\cite{PhysRevD.88.122003}. As we expand the detection frequency down to 0.1 Hz, both the signal-to-noise and the signal duration are expected to be increased for the same source mass. More importantly, this range makes it possible to search for more massive compact binary coalescences in $0.1-10$ Hz such as intermediate-mass black hole binary mergers, in addition to providing early warning for stellar-mass binary merger signals. A number of detector concepts for this mid-frequency band have been proposed. They include Superconducting Omnidirectional Gravitational Radiation Observatory (SOGRO) \cite{Paik:2016aia,Paik2020IJMPD}, DECI-hertz interferometer Gravitational-wave Observatory (DECIGO) \cite{DECIGO:2021}, , TOrsion-Bar Antenna (TOBA) \cite{TOBA:2010}, Matter wave-laser based Interferometer Gravitation Antenna (MIGA) \cite{MIGA:2018}, and Big Bang Observer (BBO) \cite{BBO:2005}.
In particular, the SOGRO concept is unique; it has maximally six detection channels measuring all components of metric perturbations, which in principle enables more efficient removal of the so-called Newtonian gravity noise (NN) that is one of the main technical challenges in the mid-frequencies~\cite{PhysRevD.92.022001}. The multi-channel detection of a single SOGRO detector in principle makes it possible to localize a GW signal.

A tunable ``free-mass'' GW detector, a predecessor of SOGRO with a single axis, was first proposed with a resonant L-C circuit~\cite{wagoner:1979}, and a wide-band resonant-mass GW detector was proposed with a resonant lever~\cite{Paik:1993a}. To improve the sensitivity, the quantum-limited SQUID~\cite{Paik:1980qz} and the inductance-bridge transducer~\cite{Paik:1986mr} were introduced. Then, a superconducting gravity gradiometer (SGG), a miniaturized GW detector with three axes, was developed for sensitive gravity measurements~\cite{Chan:1987fv} and it became the most sensitive gravity gradient sensor~\cite{Chan:1987fw,2002RScI...73.3957M}.
The original concept of SOGRO, which is essentially a large-scale version of this SGG, was proposed in Ref.~\cite{Paik:2016aia}.
Most bar-type GW detectors use the resonant amplification of a bar material responding to passing GWs, and techniques of measuring signals have evolved to using nearly quantum-limited SQUIDs at extremely low temperatures. Interferometer-type detectors, on the other hand, use freely moving test masses whose ‘relative’ motions are measured by the interference of laser lights. SOGRO also uses freely moving test masses, but its relative motions are measured by SQUID devices mounted on a rigid platform at a cryogenic temperature.

Several SOGRO designs have appeared since then. The SOGRO proposed in 2016~\cite{Paik:2016aia} has a 30 m arm length of the platform at $T=1.5$ K. The possibility of constructing higher sensitivity SOGRO has been considered by increasing the arm length to 50 m and cooling the antenna to lower temperatures: SOGRO at $T=4.2$ K and advanced SOGRO (aSOGRO) at $T=0.1$ K~\cite{2018EPJ}. Even a 100 m arm length was considered in Ref.~\cite{Paik2020IJMPD}.

Previous works presented detailed specifications of various SOGRO concepts. In this work, we study the data analysis and its target science with three different SOGRO designs: aSOGRO, SOGRO, and pSOGRO. Considering the huge differences in scales between SGG and SOGRO concepts proposed so far and the technical challenges involved, one may need a prototype project in which key technologies and their feasibility to SOGRO can be studied. Thus, we also propose a design concept of such {\it prototype} SOGRO (pSOGRO) having arm-lengths of 2 m each and test masses of 100 kg each.

In Sect.~\ref{sec:da}, various properties of the SOGRO in data analysis have been investigated, focusing on the tensorial nature. It includes the response of each channel to GWs, source localization as a single detector, and features of detecting stochastic GW backgrounds. In Sect.~\ref{sources}, the target science of SOGRO has been studied. Detection rates of binary black hole mergers and correlations for stochastic GW background are estimated for sensitivities of various SOGRO designs. In Sect.~\ref{conceptualdesign}, the conceptual design of pSOGRO and its properties are briefly described. Finally, we give conclusions with brief discussions in Sect.~\ref{discussion}.


\section{Tensorial nature of SOGRO}
\label{sec:da}

\noindent

The three-axis design of SOGRO aims to measure all GW strain tensor components~\cite{Paik:2016aia,Paik2020IJMPD}. The detector's six test masses are levitated at both ends of three orthogonal arms and move freely in arbitrary directions. The differential modes of test masses read all six components of the GW strain tensor, which implies the omnidirectional nature of SOGRO, and their common modes are rejected by superconducting circuitry~\cite{Paik:2016aia}. In more detail, in the SOGRO frequency band with $0.1-10$ Hz, the wavelengths of GWs are $(3-300) \times 10^{4}$ km. Since the distance between test masses is less than 100 m, the wave can be regarded as being uniform over the detector, e.g., $L/\lambda_{\rm GW} \le 10^{-5} \ll 1$. Then, in the laboratory frame, the coordinate position of a test mass does not move under GWs, but the proper lengths among test masses vary in time. The diagonal component of the wave $h_{xx}$, for instance, causes oscillation of the proper length between two test masses located on the $x$-axis, namely, at $+x$ and $-x$, respectively. The off-diagonal component $h_{xy}$, on the other hand, causes a scissor-like rotational oscillation of four test masses on the $x-y$ coordinate plane. Such physical motions of test masses can be combined together to measure GWs as follows~\cite{Paik:2016aia};
\begin{eqnarray}
    h_{ii} (t) &=& \frac{2}{L} \left[x_{+ii} (t)-x_{-ii} (t) \right],   \nonumber \\
h_{ij} (t) &=& \frac{1}{L} \left\{ \left[x_{+ij} (t)-x_{-ij} (t) \right]-\left[x_{-ji} (t)-x_{+ji} (t)\right] \right\} (i\not= j).
\end{eqnarray}
Here $h_{ij}$ is the GW strain tensor, $L$ the separation of two test masses on each axis in the absence of GWs, and $x_{{\pm i}j} (t)$ denotes displacement of the test mass on the $\pm i$-axis along the $j$-axis measured in the so-called Locally Lorentz (LL) frame at the center of the platform.

We note that common motions of two test masses along each coordinate axis are subtracted out, resulting in no effect on measuring the diagonal component of the wave. Only the differential motions along each coordinate axis can be measured in the diagonal channels. This feature makes the detector very insensitive to noises such as seismic noise causing common linear motions along the axis because sensors mounted on the rigid axis feel such noise as a common motion of two test masses. Similarly, differencing between two pairs of test masses in the off-diagonal channel rejects noises causing rotational common motions of four test masses on the corresponding coordinate plane.

Because SOGRO can use six detection channels, GWs from any direction can be measured, e.g., an omnidirectional detector. Note also that combining two diagonal channels, $xx$ and $yy$ channels, for instance, gives a LIGO-like interferometer detection channel. Thus, SOGRO is equivalent to a set of six small-size interferometers at the same place, three along the coordinate axes and the other three along the axes 45 degrees rotated about the coordinate axes.
In addition, scalar polarizations such as breathing and longitudinal modes of GWs, which may appear in alternative gravity theories, can be measured directly by using diagonal channels. (See Ref.~\cite{Nishizawa:2009bf} for more details.)
In SOGRO, the tiny motion of test masses induces an electrical signal which is measured by a nearly quantum-limited SQUID amplifier. To avoid the $1/f$ noise appearing in such dc SQUIDs below several kHz, a superconducting capacitor bridge transducer~\cite{Cinquegrana:1993zg} is proposed to be employed in Ref.~\cite{Paik:2016aia}. This capacitance bridge is driven at pump frequency $f_p$ well above the $1/f$ noise corner frequency of the SQUID. Then the output signal comes out at two sidebands $f_p \pm f$ with the GW frequency $f \ll f_p$.

As pointed out in Ref.~\cite{Paik2020IJMPD}, however, a terrestrial SOGRO is affected by Earth's gravity. As vertical motions of test masses are restricted by the Earth's gravity, the sensitivity for the $zz$-component of SOGRO is expected to be very poor. This results in only five components of the GW strain tensor being observable by SOGRO. Thus, these five components will be obtained only from horizontal motions of test masses.

SOGRO is capable of localizing a source as a single detector. Because SOGRO can measure each component of the GW strain tensor separately, it has practically several nonidentical GW channels. The data then can be used to find the direction toward the source. In this section, to see the omnidirectional and tensorial nature of SOGRO, we calculate the response function of a terrestrial SOGRO and briefly show the feasibility of source positioning.


\subsection{Response function}
\label{sec:response}

The tensorial data obtained by a terrestrial SOGRO is of the form
\begin{equation}
  \label{eq:data0}
  s_{ij} (t) = h_{ij} (t) + n_{ij} (t),
\end{equation}
where $n_{ij}$ is the noise. The observed data $s_{ij}$ and noise $n_{ij}$ are assumed to be symmetric as the strain tensor is.\footnote{If the observatory observes asymmetric data, then we can use a symmetrized version of Eq.~\eqref{eq:data0}, $s_{(ij)} (t) = h_{ij} (t) + n_{(ij)} (t)$, where $x_{(ij)} = \frac12 (x_{ij} + x_{ji})$ with $x = s, n$.}
Note that we will not consider $s_{33}$ in the analysis due to the gravity bias. The strain tensor in Eq.~\eqref{eq:data0} is simply written as $h_{ij} (t) = h_{ij} (t, \vec{x}=0)$ at the location of the detector, $\vec{x} = 0$.
The strain tensor can be written as
\begin{equation}
  \label{eq:hij}
  h_{ij} (t, \vec{x}) = \sum_{A=+,\times} e^A_{ij} (\hat{n}) h_A(t - \hat{n}\cdot \vec{x}/c),
\end{equation}
where $\hat{n}$ is the direction of propagation and $h_A$ the plus- and cross-polarized GW modes. The polarization tensors $e^A_{ij}$ are given by
\begin{equation}
  e^+_{ij} = \hat{u}_i \hat{u}_j - \hat{v}_i \hat{v}_j, \qquad e^\times_{ij} = \hat{u}_i \hat{v}_j + \hat{v}_i \hat{u}_j,
\end{equation}
where two unit vectors $\hat{u}$ and $\hat{v}$ are orthogonal to $\hat{n}$ and to each other.
These three unit vectors are explicitly given in terms of the polar and azimuthal angles $(\theta, \phi)$ of the propagating direction $\hat{n}$, respectively, and the polarization angle $\psi$ as in Eq.~\ref{unit-vectors} in Appendix A.
We choose the polarization angle $\psi = 0$ for convenience in what follows unless it is explicitly written.
Note that, in this convention, the direction of $\hat{u}$ with respect to which the polarizations are defined is determined once the propagation direction $\hat{n}$ is given, as the intersection between the perpendicular plane to $\hat{n}$ and the $x$\nobreakdash-$y$~plane at a detector.

\begin{table}[tp]

\caption{\label{table:response} Responses of each channel in the SOGRO detector for plus ($+$), cross ($\times$), $x$, $y$, breathing ($b$), and longitudinal ($\ell$) polarization modes. The $+$ and $\times$ polarizations correspond to tensor modes, the $x$ and $y$ are vector modes, and the $b$ and $\ell$ polarizations are relevant to scalar modes. The responses of the (22) and the (31) channels are the same with the (11) and the (23) channels, respectively, up to $\pi/2$ rotation along the $z$-axis. The (33) channel is not shown as this is not used in sensitivity calculation. Combined (total) responses of all 5 channels are also depicted. The total responses show the omnidirectional nature of SOGRO; however, the total response of a terrestrial SOGRO to the $\ell$ mode vanishes in the direction of $z$-axis. In the last column, the general responses of a laser interferometer such as LIGO, Virgo or KAGRA are presented for comparison.}

  \centering
  \renewcommand{\arraystretch}{0.5}
  \begin{tabular}{
    >{\centering\arraybackslash}m{.8in}
    >{\centering\arraybackslash}m{.12\textwidth}
    >{\centering\arraybackslash}m{.12\textwidth}
    >{\centering\arraybackslash}m{.12\textwidth}
    >{\centering\arraybackslash}m{.12\textwidth}
    >{\centering\arraybackslash}m{.12\textwidth}
  }
  \toprule
    & \multicolumn{4}{c}{SOGRO} & Interferometer \\
  \cmidrule(lr){2-5}
    Modes & (11) & (12) & (23) & total & \\
  \midrule
  \rule{0pt}{0.5\normalbaselineskip} 
    \(+\) & \includegraphics[width=.13\textwidth]{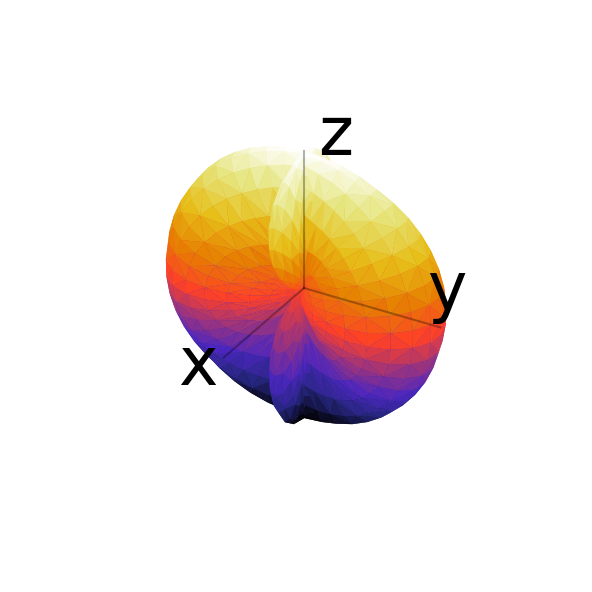} & \includegraphics[width=.13\textwidth]{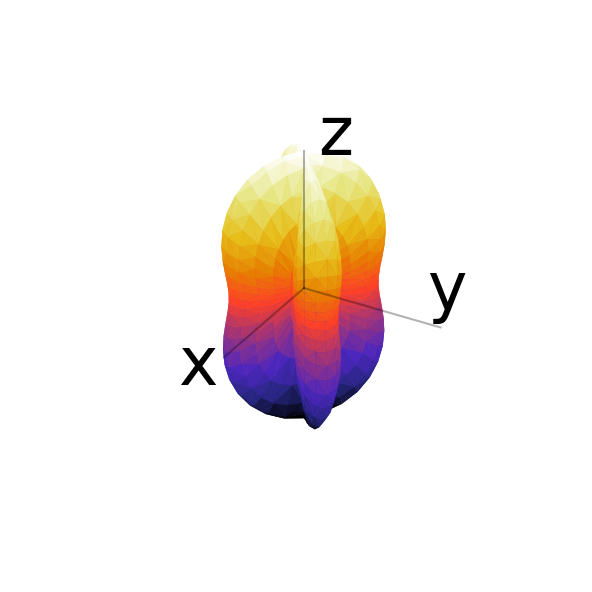} & \includegraphics[width=.13\textwidth]{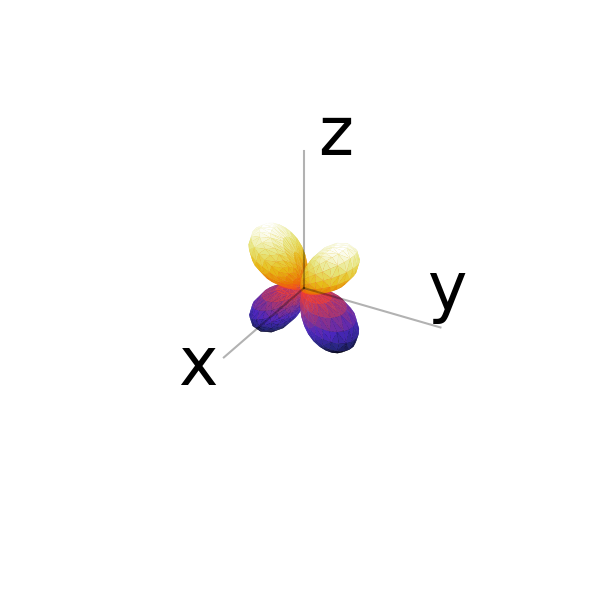} & \includegraphics[width=.13\textwidth]{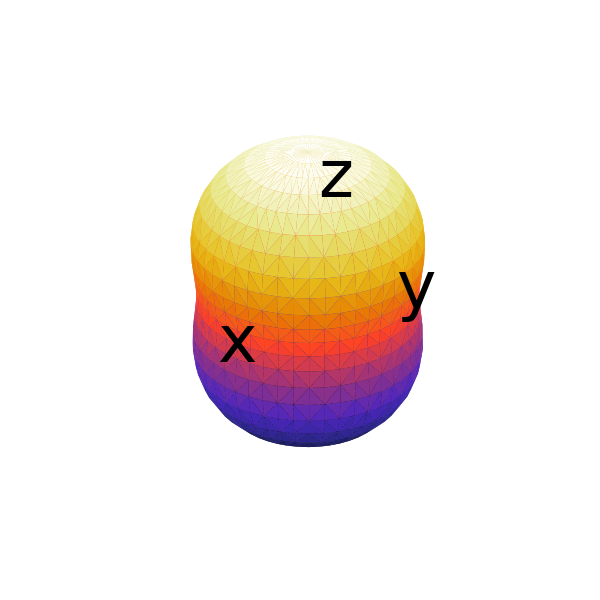} & \includegraphics[width=.13\textwidth]{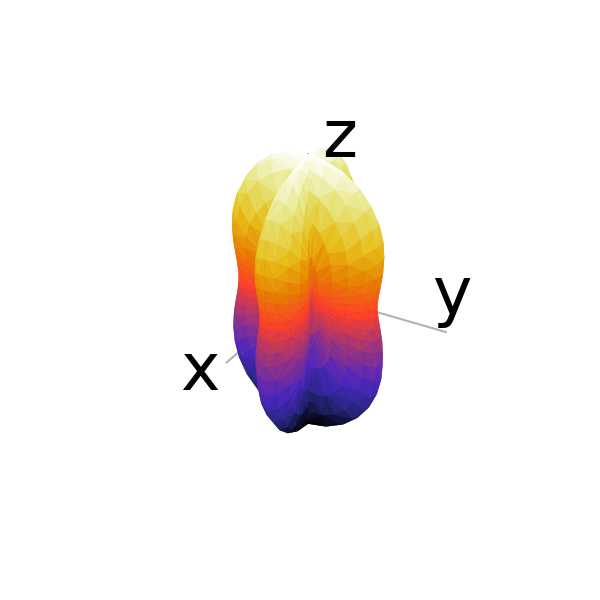} \\[-15pt]

    \(\times\) & \includegraphics[width=.13\textwidth]{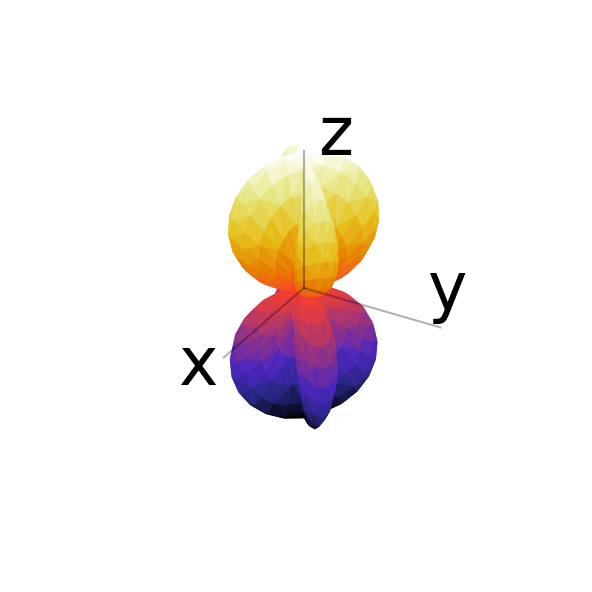} & \includegraphics[width=.13\textwidth]{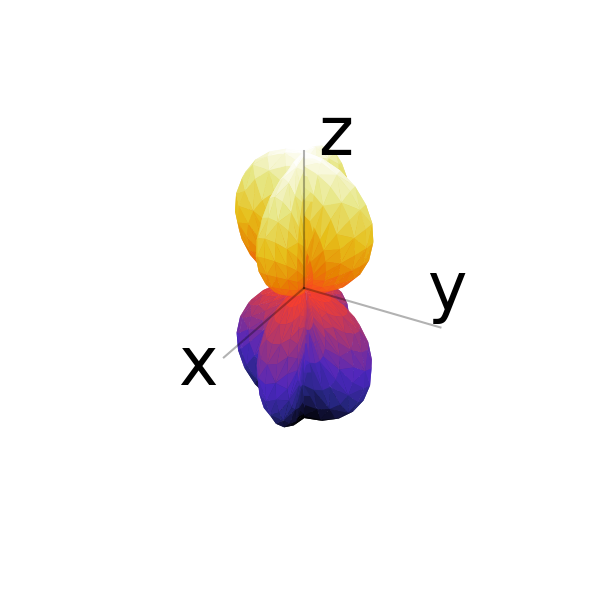} & \includegraphics[width=.13\textwidth]{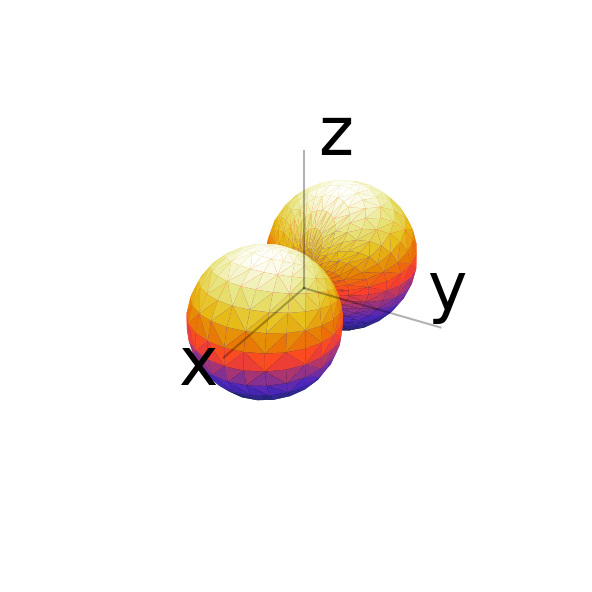} & \includegraphics[width=.13\textwidth]{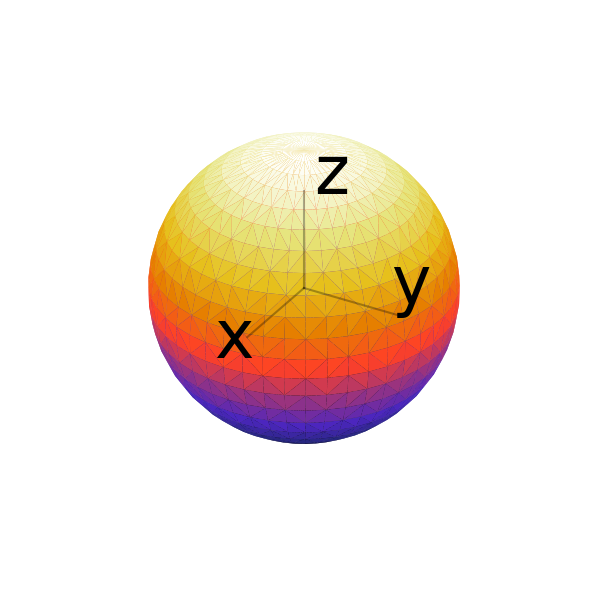} & \includegraphics[width=.13\textwidth]{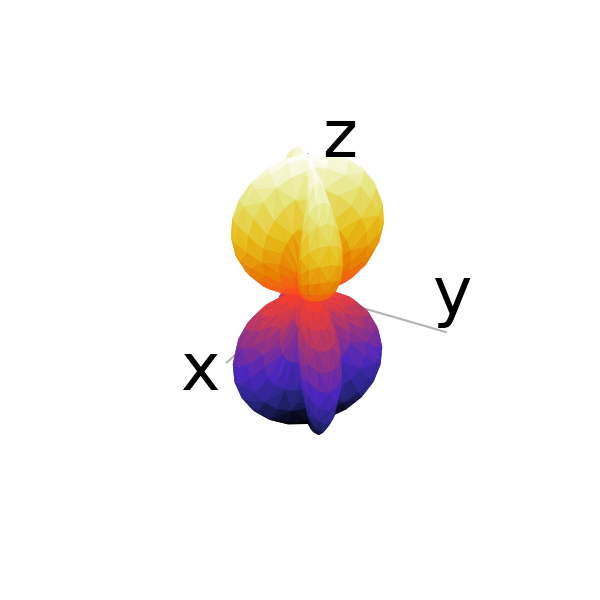} \\[-15pt]

    \(\textrm{unpolarized} \newline \textrm{tensor}\) & \includegraphics[width=.13\textwidth]{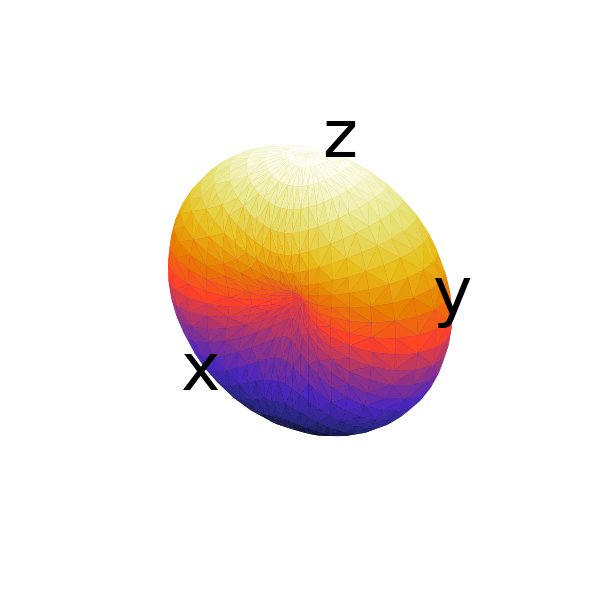} & \includegraphics[width=.13\textwidth]{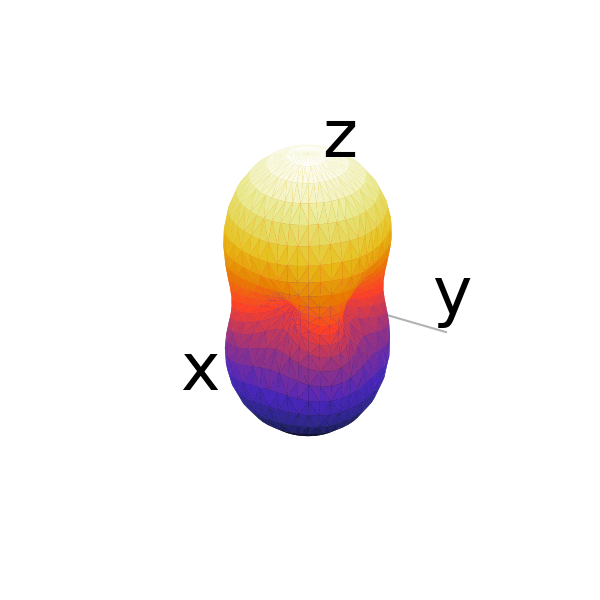} & \includegraphics[width=.13\textwidth]{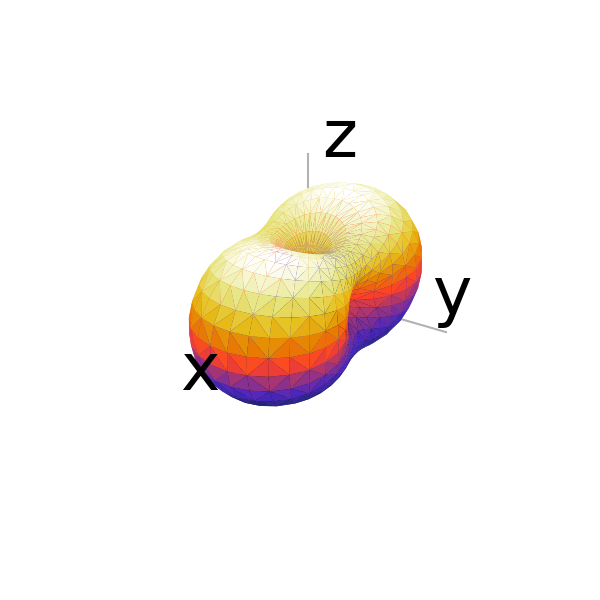} & \includegraphics[width=.13\textwidth]{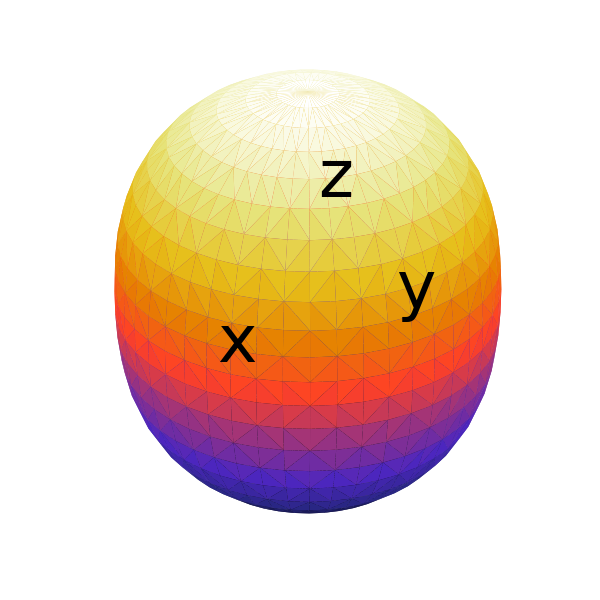} & \includegraphics[width=.13\textwidth]{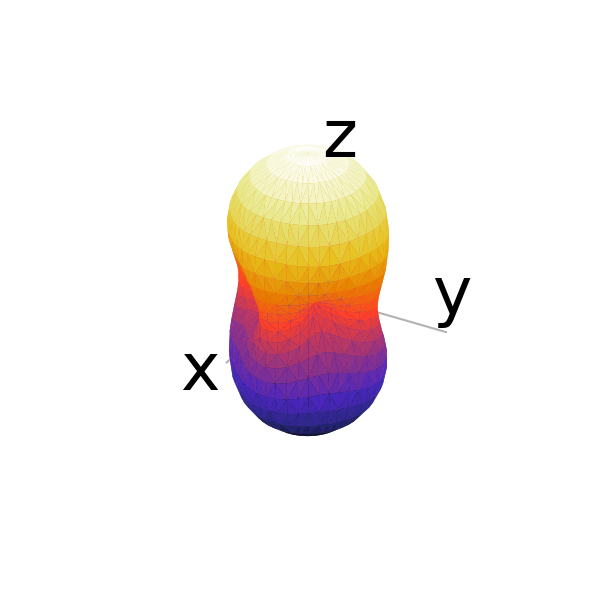} \\
    [-13pt]
    
  \midrule
    \(x\) & \includegraphics[width=.13\textwidth]{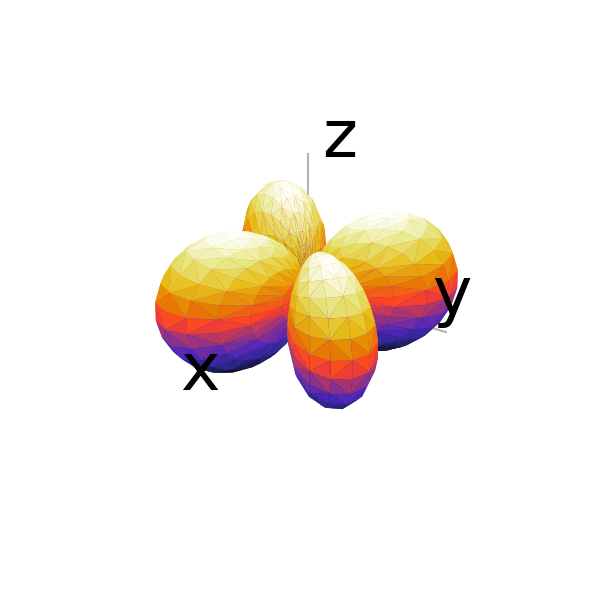} & \includegraphics[width=.13\textwidth]{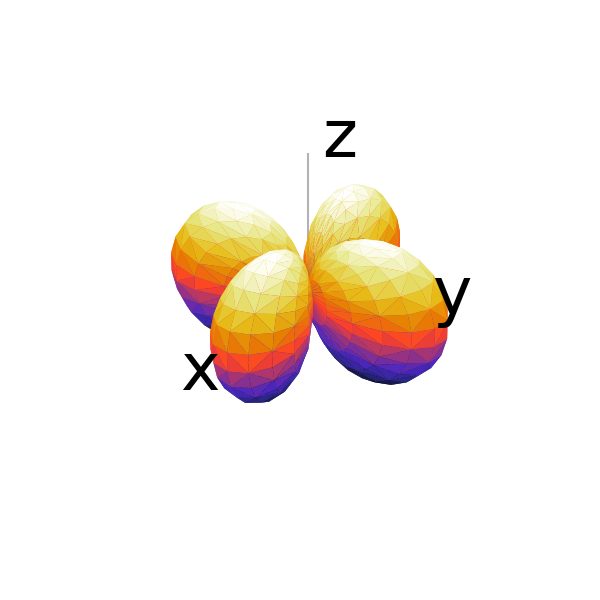} & \includegraphics[width=.13\textwidth]{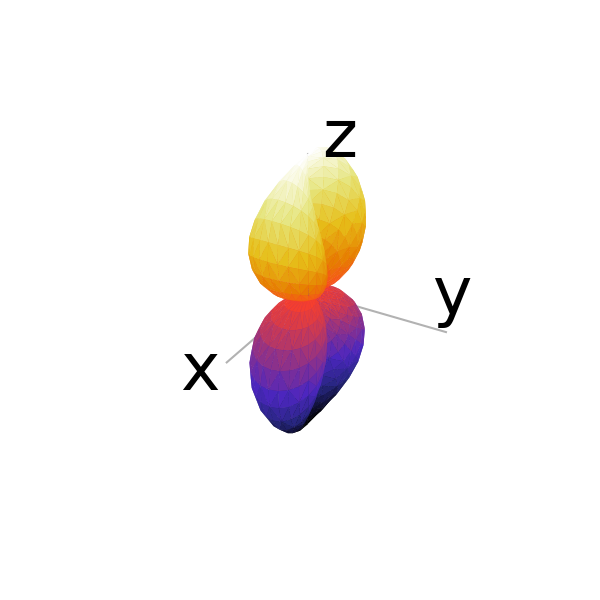} & \includegraphics[width=.13\textwidth]{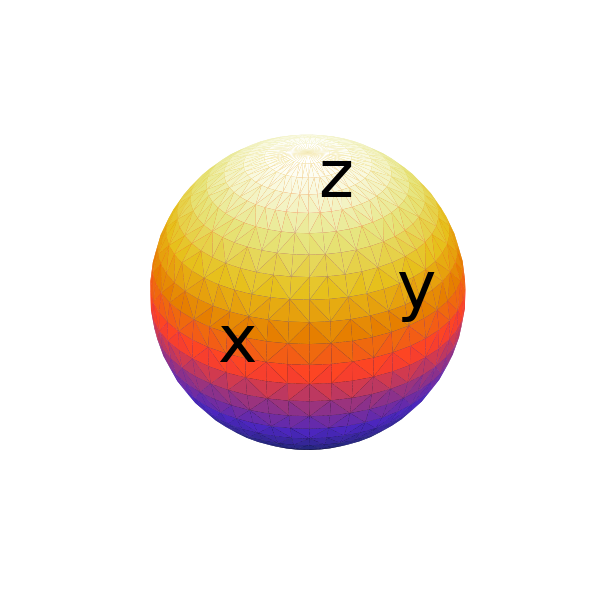} & \includegraphics[width=.13\textwidth]{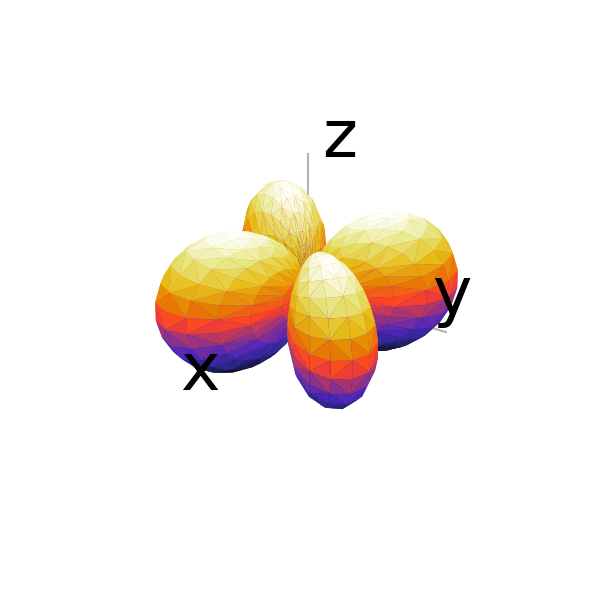} \\[-15pt]    

    \(y\) & \includegraphics[width=.13\textwidth]{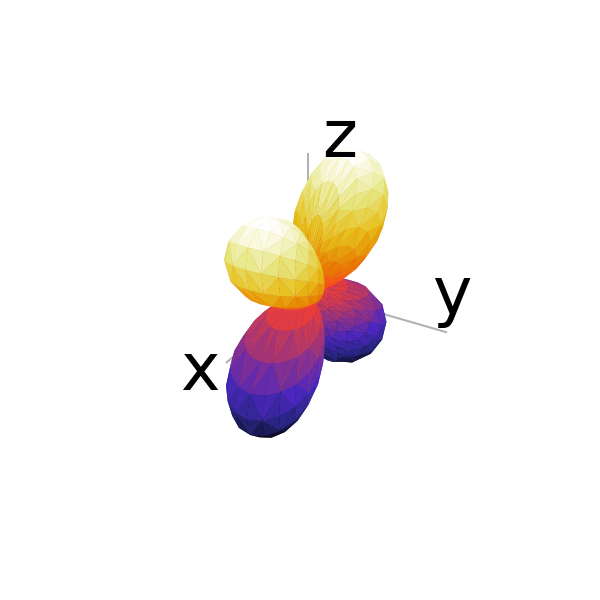} & \includegraphics[width=.13\textwidth]{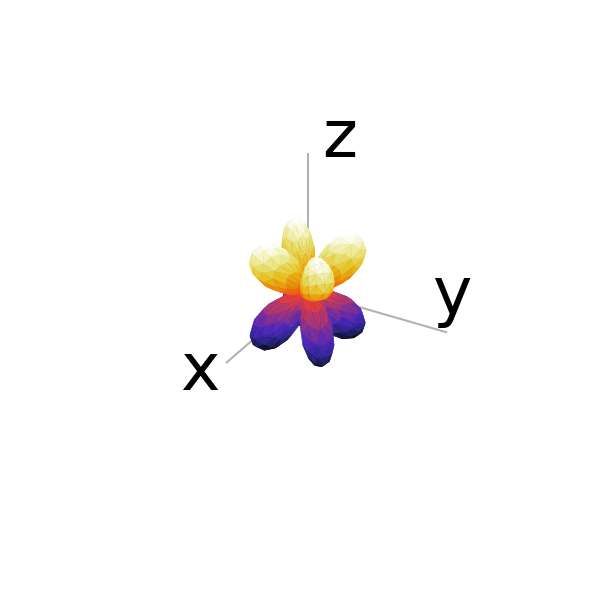} & \includegraphics[width=.13\textwidth]{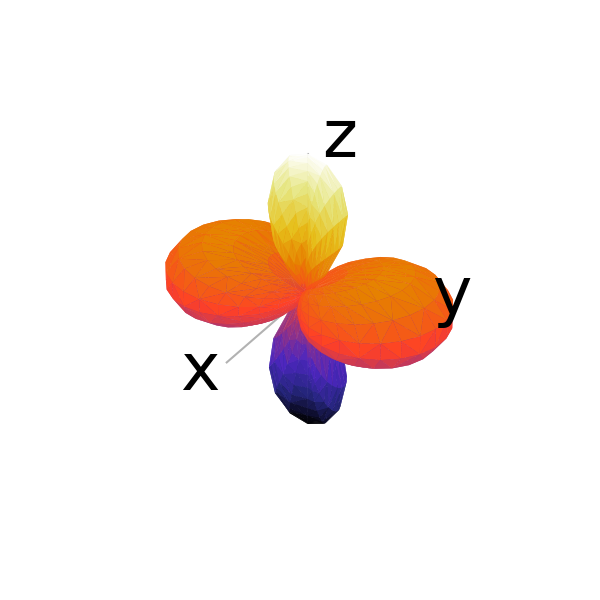} & \includegraphics[width=.13\textwidth]{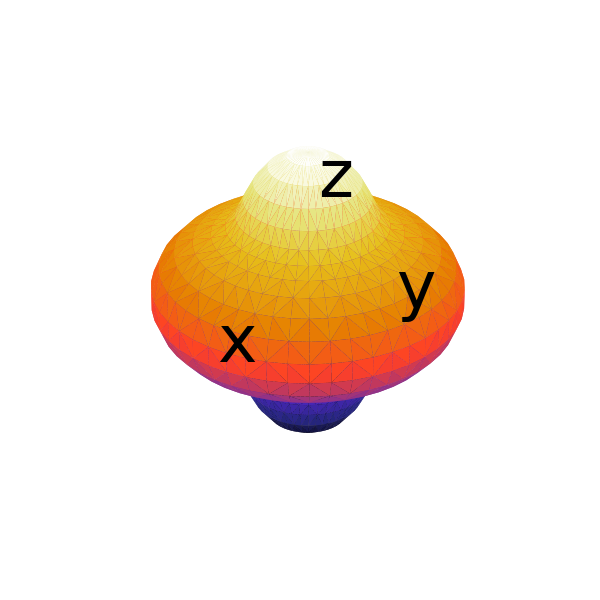} & \includegraphics[width=.13\textwidth]{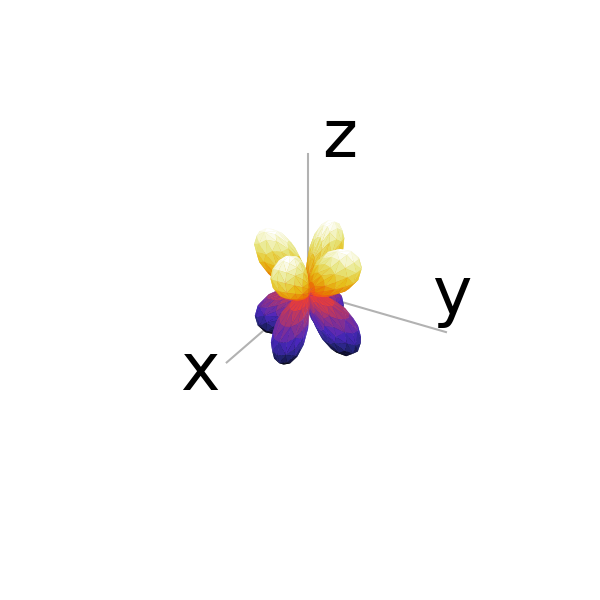} \\[-15pt]    

    \(\textrm{unpolarized} \newline \textrm{vector}\) & \includegraphics[width=.13\textwidth]{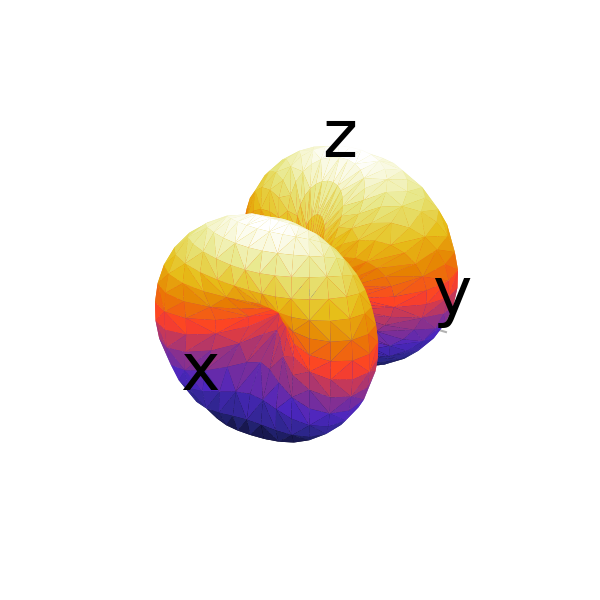} & \includegraphics[width=.13\textwidth]{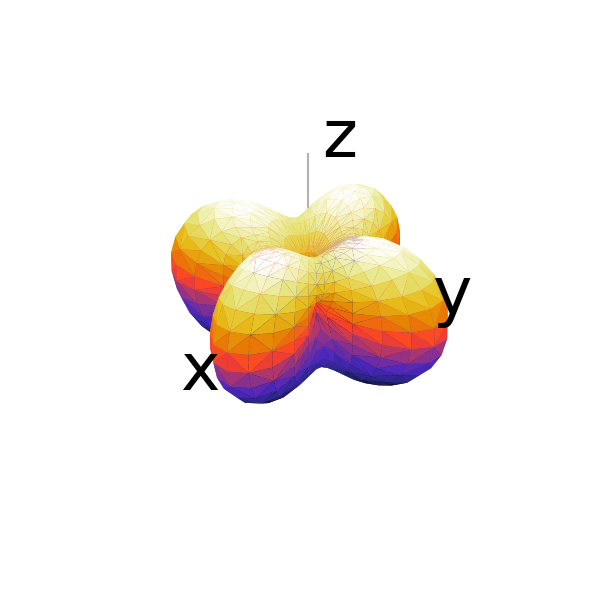} & \includegraphics[width=.13\textwidth]{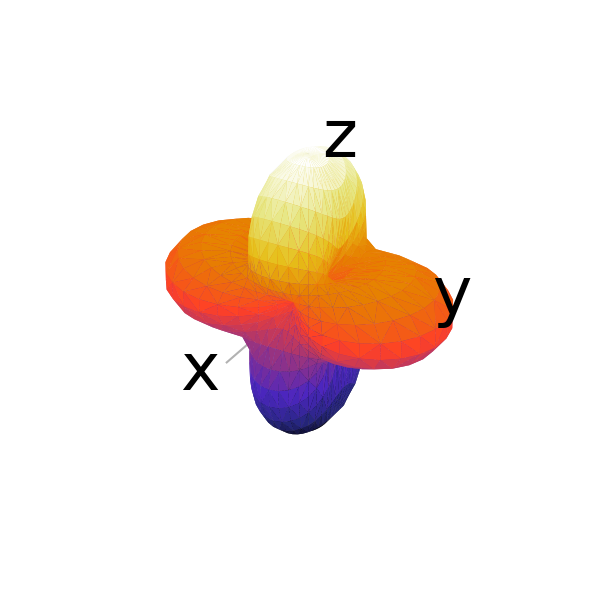} & \includegraphics[width=.13\textwidth]{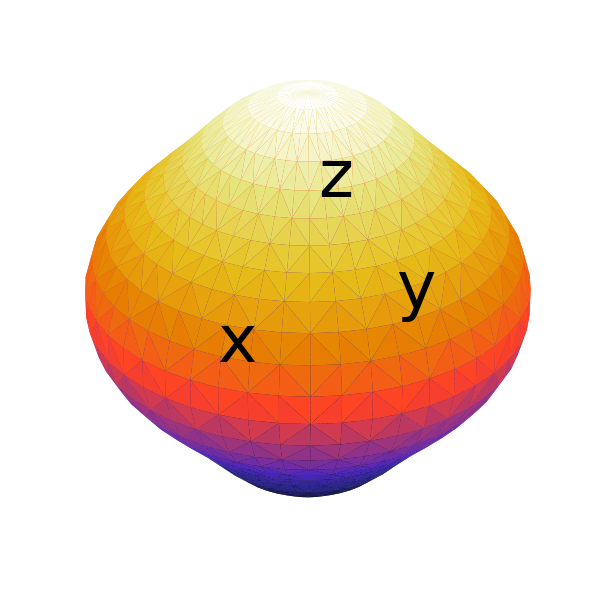} & \includegraphics[width=.13\textwidth]{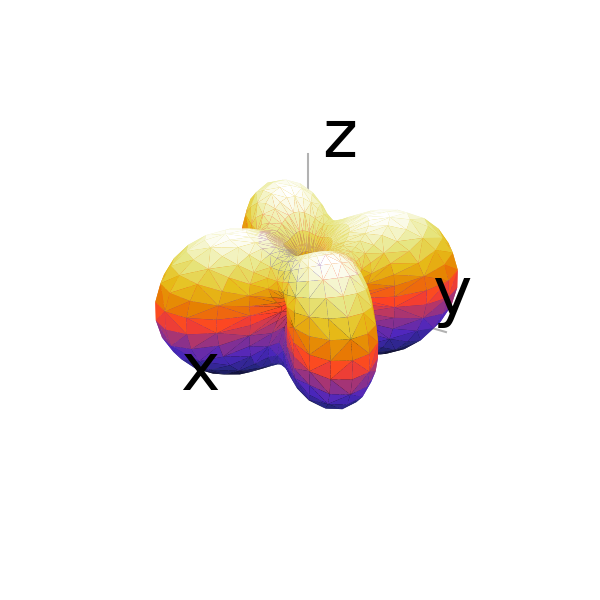} \\[-13pt]
    
  \midrule
    \(b\) & \includegraphics[width=.13\textwidth]{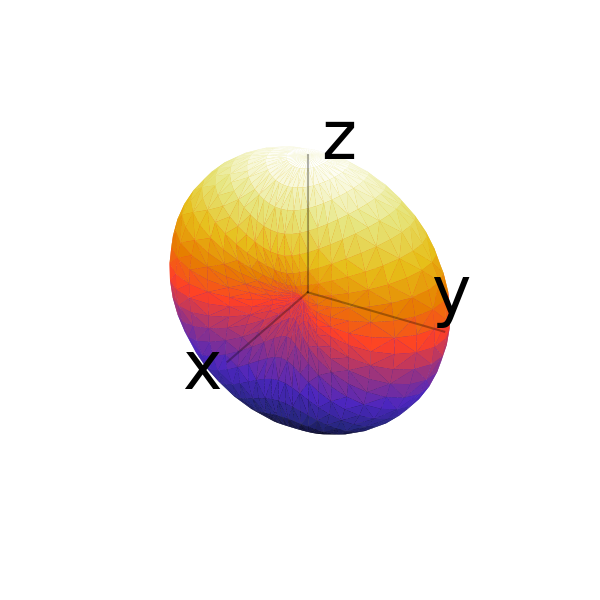} & \includegraphics[width=.13\textwidth]{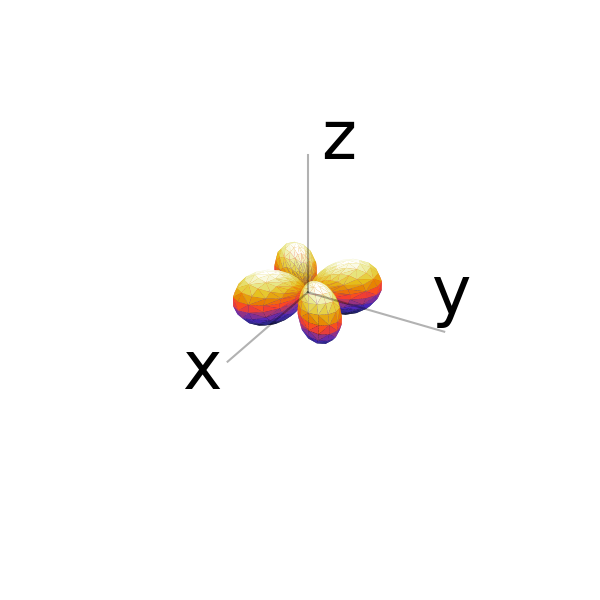} & \includegraphics[width=.13\textwidth]{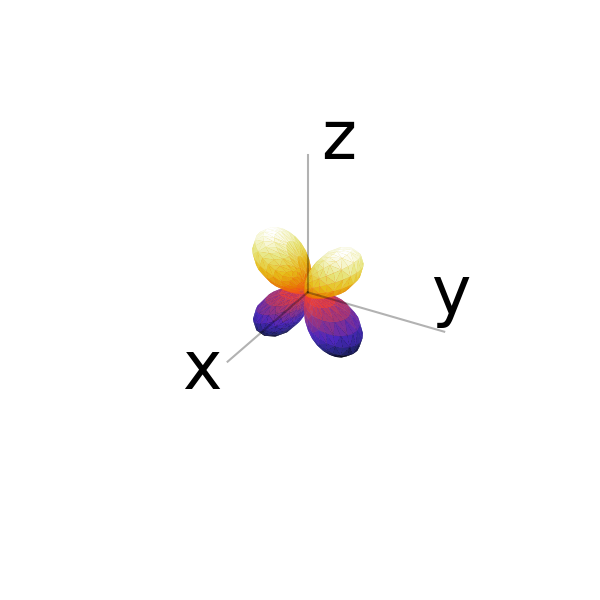} & \includegraphics[width=.13\textwidth]{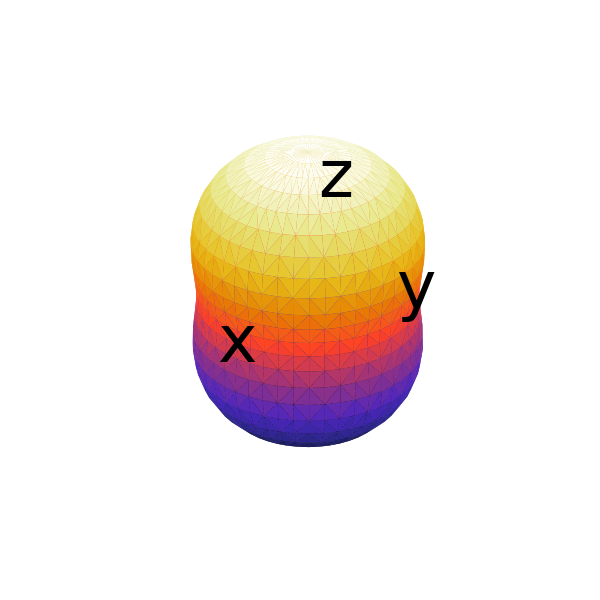} & \includegraphics[width=.13\textwidth]{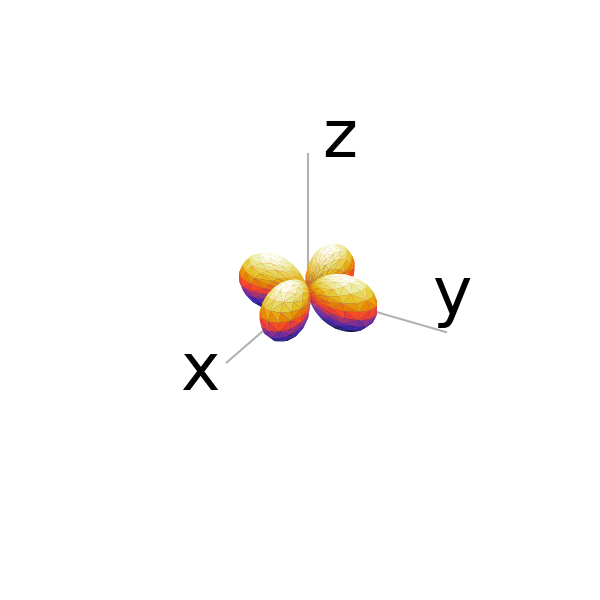} \\[-15pt]    

    \(\ell\) & \includegraphics[width=.13\textwidth]{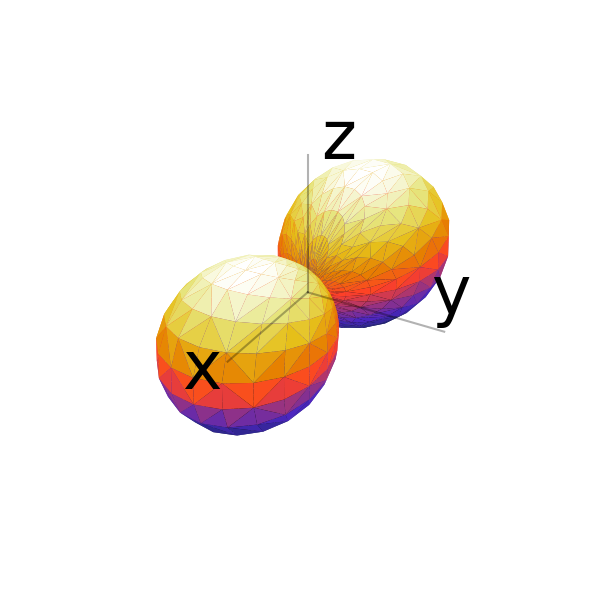} & \includegraphics[width=.13\textwidth]{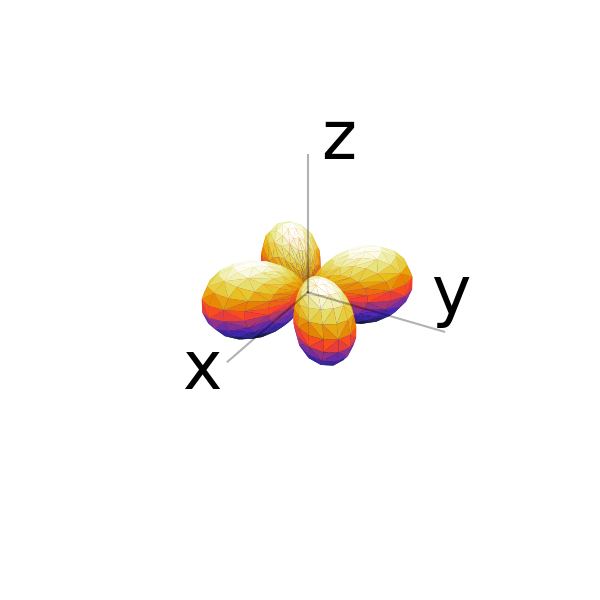} & \includegraphics[width=.13\textwidth]{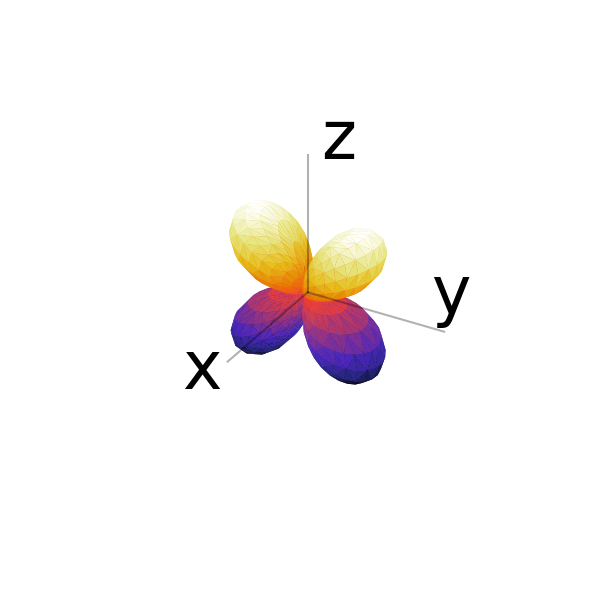} & \includegraphics[width=.13\textwidth]{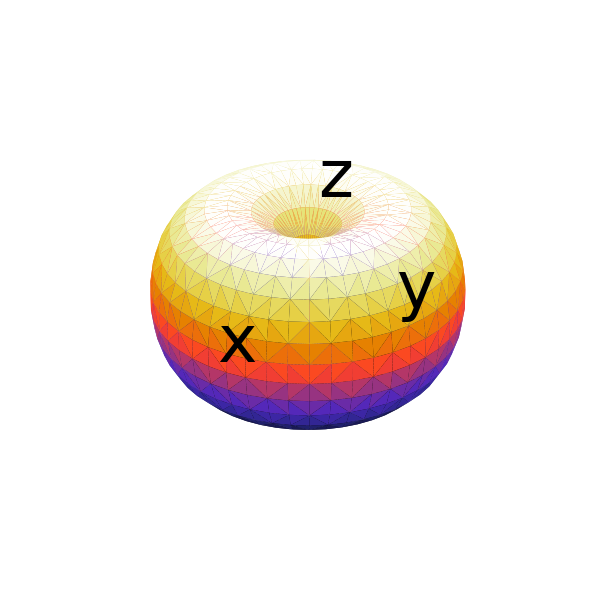} & \includegraphics[width=.13\textwidth]{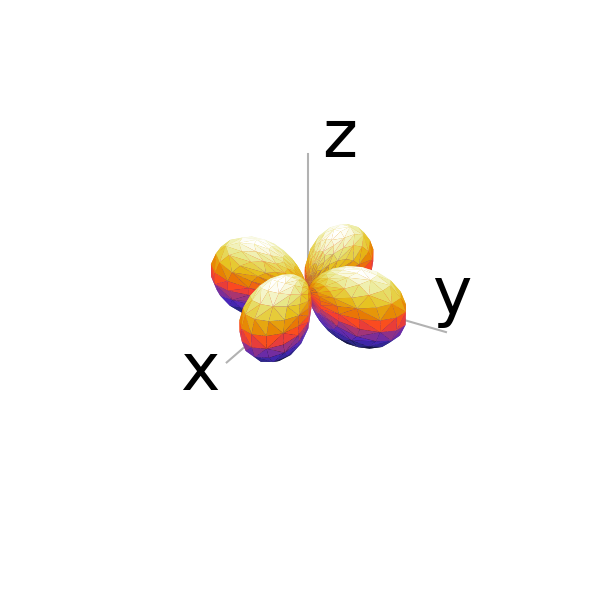} \\[-15pt]        
  \bottomrule
  \end{tabular}
  \renewcommand{\arraystretch}{1}
\end{table}

Recall that the laser interferometer observes scalar data $s(t) = D^{ij} h_{ij} (t) + n(t)$ with $D^{ij} = \frac12 \left( \hat{x}^i \hat{x}^j - \hat{y}^i \hat{y}^j \right)$, where $\left( \hat{x}, \hat{y}, \hat{z} \right) = R_z(-\psi) R_x(\theta) R_z(\pi/2 - \phi) \left( \hat{u}, \hat{v}, \hat{n} \right)$ are the basis vectors in the detector frame and the two arms of the laser interferometer are located in the directions of $x$ and $y$ axes. Then, the response of the laser interferometer is given by
$F_A = D^{ij} e^A_{ij} = \frac12 \left(e^A_{11} - e^A_{22}\right)$ for the polarization $A$.
Explicitly, the full response functions are given by\footnote{
At first sight, these formulae look slightly different from those in literature; \textit{e.g.}, they are different from those in Refs.~\cite{Christensen:1992wi,Nishizawa:2009bf} in the overall sign. In addition, those in Refs.~\cite{Sathyaprakash:2009xs,Varma:2014jxa} agree with $(-F_+)$ and $F_\times$ in this paper (equivalently, $F_+$ and $(-F_\times)$ in Refs.~\cite{Christensen:1992wi,Nishizawa:2009bf}). These mismatches come from the differences in the coordinate system in the source frame, and all the formulae in this paper and the literature agree with each other up to some rotations. For example, we use the \textit{z-x-z} rotation to obtain the coordinate system in the source frame, while the \textit{z-y-z} rotation is used in Ref.~\cite{Nishizawa:2009bf}. The \textit{z-x-z} rotation is used in Ref.~\cite{Christensen:1992wi} but $\hat{n}$ is rotated by $(-\pi/2)$ with respect to $z$-axis, and so on.
}
\begin{eqnarray}
    F_+ (\theta,\phi,\psi) &=& -\frac{1}{2} \left( 1+\cos^2\theta \right) \cos 2\phi \cos 2\psi + \cos \theta \sin 2\phi \sin 2\psi  \\
    F_{\times} (\theta,\phi,\psi) &=& \frac{1}{2} \left( 1+\cos^2\theta \right) \cos 2\phi \sin 2\psi +\cos \theta \sin 2\phi \cos 2\psi.
\end{eqnarray}
Then the signal observed at the laser interferometer, ignoring the noise, becomes
\begin{equation}
    h(t) = D^{ij}h_{ij}(t) = F_+(\theta,\phi,\psi) h_+(t) +F_{\times}(\theta,\phi,\psi) h_{\times}(t).
\end{equation}
Note that the polarization angle can be set to be $\psi=0$ without loss of generality. In order to fully determine the four unknowns ({\it e.g.}, two direction angles and two polarization amplitudes of the GW), therefore, one needs at least three different interferometers from which three outputs $h(t)$ and two independent time delays can be obtained.

Since a SOGRO observes tensorial data (Eq.~\ref{eq:data0}), the response of each component of the data is simply given by $|e^A_{ij}|$ for polarization $A$ and the total response of a SOGRO is obtained as a quadrature sum of responses from each component. The response of a SOGRO is depicted in Table~\ref{table:response}.
The response of the diagonal components --- $h_{11}$ and $h_{22}$ --- is the same as the response of the cylindrical bar detector up to some rotations.
Note that the SOGRO has multiple detection channels. Thus, the signal data measured at each detection channel can be written as
\begin{equation}
h_C(t)=D^{ij}_C h_{ij} (t) = \sum_{A} D^{ij}_C e^A_{ij} h_A(t) = \sum_{A} F^C_A(\theta,\phi,\psi) h_A(t).
\label{SOGROoutput}
\end{equation}
Here $D^{ij}_C$ are the detector tensors for the detection channels $C=(11), (22), (12), (23), (31)$, respectively. Namely, the channel $C=(11)$ implies the detection for differential motions of the test masses at the ends of the $x-$axis of the SOGRO, and so $D^{ij}_{11}=\hat{x}^i \hat{x}^j=\delta^i_1\delta^j_1$. Similarly, $D^{ij}_{22}=\hat{y}^i\hat{y}^j, D^{ij}_{12}=\frac{1}{2}\left( \hat{x}^i\hat{y}^j+\hat{y}^i\hat{x}^j \right), D^{ij}_{23}=\frac{1}{2}\left( \hat{y}^i\hat{z}^j+\hat{z}^i\hat{y}^j \right), D^{ij}_{31}=\frac{1}{2}\left( \hat{z}^i\hat{x}^j+\hat{x}^i\hat{z}^j \right)$.
Then the response functions $F^C_A(\theta,\phi,\psi)=D_C^{ij}e^A_{ij}=e^A_C$ become
\begin{eqnarray}
    F^C_+ (\hat{\bf{n}}; \psi) &=& F^C_+ (\hat{\bf{n}}) \cos 2\psi + F^C_{\times} (\hat{\bf{n}}) \sin 2\psi  \nonumber \\
    F^C_{\times} (\hat{\bf{n}}; \psi) &=& -F^C_+ (\hat{\bf{n}}) \sin 2\psi +F^C_{\times} (\hat{\bf{n}}) \cos 2\psi.
\label{SOGROresponsepm}
\end{eqnarray}
Here $F^C_A (\hat{\bf{n}}) \equiv F^C_A(\theta,\phi,\psi=0)$ are given by 
\begin{eqnarray}
    F^{(11)}_+(\hat{\bf{n}}) &=& 1-\left( 1+\cos^2\theta \right) \cos^2\phi , \quad
    F^{(11)}_{\times}(\hat{\bf{n}}) = \cos \theta \sin 2\phi ,  \nonumber  \\
    F^{(22)}_+(\hat{\bf{n}}) &=& 1-\left( 1+\cos^2\theta \right) \sin^2\phi , \quad
    F^{(22)}_{\times}(\hat{\bf{n}}) = -\cos \theta \sin 2\phi ,  \nonumber  \\
    F^{(12)}_+(\hat{\bf{n}}) &=& -\frac{1}{2}\left( 1+\cos^2\theta \right) \sin 2\phi , \quad
    F^{(12)}_{\times}(\hat{\bf{n}}) = -\cos\theta \cos2\phi ,  \nonumber  \\
    F^{(23)}_+(\hat{\bf{n}}) &=& \frac{1}{2}\sin2\theta \sin\phi , \quad
    F^{(23)}_{\times}(\hat{\bf{n}}) = \sin\theta \cos\phi ,  \nonumber  \\
    F^{(31)}_+(\hat{\bf{n}}) &=& \frac{1}{2}\sin2\theta \cos\phi , \quad
    F^{(31)}_{\times}(\hat{\bf{n}}) = -\sin\theta \sin\phi .
\end{eqnarray}
The magnitudes of some of them are shown in Table~\ref{table:response}. 
Note that $F_A^C$ and $h_C$ can be simply written as $F^{C=(ij)}_A = e^A_{C=(ij)} = e^A_{ij}$ and $h_{C=(ij)} = h_{ij}$ with the help of the symmetry in $e^A_{ij}$.

Notice that, in principle, a single SOGRO detector can determine all four unknowns because its single measurement with five channels could produce five outputs as in Eq.~(\ref{SOGROoutput}). Note also that the off-axis channel of SOGRO essentially has the same response as LIGO's, {\it e.g.,} $F^{(12)}_A(\theta,\phi+\pi/4,\psi)=F^{LIGO}_A(\theta,\phi,\psi)$. It follows because the motion of end masses in scissor modes is equivalent to the oscillation of mirrors in either $x-$ or $y-$ arm in the interferometer. Putting this in another way, one can see that the rotation of the laser interferometer by $\pi/4$ with respect to the $z-$axis yields its detector tensor ${D^{ij}}'=R_z(\pi/4) \frac12 \left( \hat{x}^i \hat{x}^j - \hat{y}^i \hat{y}^j \right) R_z^{-1}(\pi/4) = \frac12 \left( \hat{x}^i \hat{y}^j + \hat{y}^i \hat{x}^j \right) = D^{ij}_{(12)}$. Similarly, the SOGRO channels $(23)$ and $(32)$ with appropriate rotations and the combination of SOGRO channels such as $\frac{1}{2}\left( D^{ij}_{(11)} -D^{ij}_{(22)} \right)$ can produce the response identical to that of the laser interferometer.
It should be pointed out that these antenna patterns of five detection channels in total make the SOGRO being omnidirectional as can be seen by the total response function at the column next to the last in Table~\ref{table:response}. Namely, at least some of the five channels can measure a non-vanishing signal regardless of the source direction and polarization of a GW.
This omnidirectional nature of SOGRO is similar to the spherical resonance detectors such as TIGA and mini-GRAIL~\cite{spherical,minigrailweb}, but SOGRO can cover a much broad frequency band of $0.1-10$ Hz. 

In alternative gravity theories, in addition to the plus- and cross-polarizations in general relativity, there exist more polarization degrees of freedom such as vector ($x$ and $y$), breathing ($b$) and longitudinal ($\ell$) polarizations~\cite{Nishizawa:2009bf,Hyun:2018pgn}. The polarization tensors for the vector and scalar modes are given by $e^x_{ij} = \hat{u}_i \hat{n}_j + \hat{n}_i \hat{u}_j$, $e^y_{ij} = \hat{v}_i \hat{n}_j + \hat{n}_i \hat{v}_j$, $e^b_{ij} = \hat{u}_i \hat{u}_j + \hat{v}_i \hat{v}_j$ and $e^\ell_{ij} = \sqrt{2} \hat{n}_i \hat{n}_j$. The response functions of the laser interferometer for GWs having such polarizations are given by Eqs.~(\ref{interferometer-vbl-modes}) in Appendix A.
For SOGRO channels, on the other hand, they are given by Eqs.~(\ref{SOGRO-vbl-modes}) in Appendix A.
One can easily see that detections at the laser interferometers cannot distinguish the breathing and the longitudinal modes whereas they are not degenerate at the SOGRO detections. In the presence of general polarizations, however, at least two SOGRO detectors at different sites are necessary
because we have four more unknowns associated with the amplitudes of four extra polarizations.


\subsection{Combined noise spectral density}
\label{nsd}

\noindent
A single SOGRO can be considered as a network of detectors because each channel plays the role of a GW detector, which has its own noise spectral density (NSD). To find a representative NSD of a SOGRO as a single GW antenna, in this section, we try to combine NSDs of channels in a SOGRO. 

Firstly, we recall that the optimal matched filter signal-to-noise ratio (SNR) for the interferometer detector such as LIGO, Virgo or KAGRA is given by
\begin{align}
  \label{eq:ligo:matchedfilter}
  \rho^2 &= 4 \int_{0}^{\infty} df\, \frac{\left| \tilde{h}(f) \right|^2}{S_n (f)} \notag \\
  &= \left\{ 
  \begin{aligned}
    & 2 \int_{0}^{\infty} df\, \frac{\left| \tilde{h}_+(f) \right|^2 + \left| \tilde{h}_\times(f) \right|^2}{S_n (f)}, &&\text{(optimal direction)} \\
    & \frac{4}{5} \int_{0}^{\infty} df\, \frac{\left| \tilde{h}_+(f) \right|^2 +  \left| \tilde{h}_\times(f) \right|^2}{S_n (f)}, &&\text{(random direction)}
  \end{aligned} \right.
\end{align}
where $S_n (f)$ is the single-sided NSD of the interferometer.
The optimal direction in Eq.~\eqref{eq:ligo:matchedfilter} represents that the GW under consideration is coming from the $z$-direction ({\it e.g.,} the vertical direction to the interferometer plane) with a random polarization, where the antenna response of the interferometer becomes maximum. The expected SNR for the optimal direction is calculated by adopting the polarization averaging.
The random direction in Eq.~\eqref{eq:ligo:matchedfilter}, on the other hand, represents that the GW is coming from a random direction with a random polarization. Hence, the expected SNR for the random direction is calculated by adopting the total angle averaging. Then, one can define two different versions of the NSD for the interferometer, namely, the \textit{optimal} NSD, $S_n$, and the \textit{angle-averaged} NSD, $(5/2)S_n$. These NSDs are used to define the sensitivities of the interferometer GW detector.

For the SOGRO detector having multiple channels, if we assume that the noise $n_{ij}$ is stationary Gaussian and each component is uncorrelated with the others, then we have
the combined matched filter SNR~\cite{Finn:2000hj},
\begin{eqnarray}
  \label{eq:snr:matchedfilter}
  \rho^2 &=& \sum_C \left[ 4 \int_{0}^{\infty} df\, \frac{\left| \tilde{h}_C (f) \right|^2}{S_{n,C} (f)} \right] \nonumber \\
  &=& \sum_C \left[ 4 \int_{0}^{\infty} df\, \frac{F_{+,C}^2 \left| \tilde{h}_+ (f) \right|^2 + F_{\times,C}^2 \left| \tilde{h}_\times (f) \right|^2}{S_{n,C} (f)} \right],
\end{eqnarray}
where $S_{n,C}$ is the single-sided NSD of the channel $C$. Note that, considering GWs with identical waveforms coming in different directions, the combined SNRs in Eq.~\eqref{eq:snr:matchedfilter} are not the same in general. It depends on the direction of the GW source. The combined SNR is expected to have its maximal value when a GW is coming from the $z$-direction, because the maximum response occurrs at the $z$-direction as seen in Table~\ref{table:response}.

Now, one can define two kinds of combined NSDs as in the case of interferometer: one obtained from the maximum SNR and the other from the angle-averaged SNR. Namely, for the maximum SNR, the optimal NSD will be defined as
\begin{equation}
    \label{nsd:optimal}
    S_n^\mathrm{(optimal)} = \left[ S_{n,(11)}^{-1} + S_{n,(22)}^{-1} + S_{n,(12)}^{-1} \right]^{-1}.
\end{equation}
This optimal NSD will be used to denote the sensitivities ({\it i.e.,} $\sqrt{S_n^\mathrm{(optimal)}}$) for different SOGRO designs in Sect.~\ref{sources}.
Next, for the angle-averaged SNR, the angle-averaged NSD is defined as
\begin{equation}
    \label{nsd:averaged}
    S_n^\mathrm{(averaged)} = \left[ \frac{8}{15} \sum_{C\in\mathcal{D}} S_{n,C}^{-1} + \frac{2}{5} \sum_{C\in\mathcal{D}'} S_{n,C}^{-1} \right]^{-1},
\end{equation}
where $\mathcal{D} = \{\textrm{diagonal channels}\}$ and $\mathcal{D}' = \{\textrm{off-diagonal channels}\}$. For a terrestrial SOGRO, assuming that $S_{n,(11)} = S_{n,(22)} \approx 2S_{n,(12)} \approx S_{n,(23)} = S_{n,(31)}$, we have $S_n^\mathrm{(optimal)} \approx \frac12 S_{n,(12)} \approx (2/3) S_n^\mathrm{(averaged)}$.


\subsection{Source localization}

In this section, we show a simple visualization of the source positioning by a single SOGRO. For the accurate source localization, we may adopt the localization algorithms for the LIGO-Virgo-KAGRA network (see, e.g., Ref.~\cite{Abbott:2020qfu} and references therein.) because a SOGRO, having multiple channels, can be regarded as a network of detectors. This is out of the scope of the present paper though. Obviously, the triangulation cannot be used for a single SOGRO; however, the response functions of all channels guide us to obtain the position of the source.

Since we observe the tensorial data of Eq.~\eqref{eq:data0}, we can find the position of the source even with a single SOGRO.
Considering a simulated GW signal propagating in the direction $\hat{n}_0 = (\theta_0, \phi_0)$, the position of the source is in the opposite direction, $-\hat{n}_0$.
In order to find $\hat{n}_0$, we first start with the signal $h_{ij}$ without noise for simplicity.
Eq.~\eqref{eq:hij} can be rewritten as
\begin{equation}
  \left[
  \begin{array}{ccc}
    h_+ & h_\times & 0 \\
    h_\times & -h_+ & 0 \\
    0 & 0 & 0
  \end{array}
  \right]
  = R_x(\theta) R_z(\pi/2 - \phi) [ h_{ij} ] R_z(\phi - \pi/2) R_x(-\theta),
\end{equation}
where $R_{k}$ is the rotation matrix with respect to $k$-axis.
Since $h_{ij}$ is symmetric and traceless and its determinant vanishes, we have 4 linearly independent equations:
\begin{align}
  \label{eq:loc:hp}
  & h_+ = h_{11} \sin^2 \phi - h_{12} \sin (2\phi) + h_{22} \cos^2 \phi, \\
  \label{eq:loc:hc}
  & h_\times = (h_{23} \cos \phi - h_{13} \sin \phi) \sin \theta \nonumber \\
  & \qquad + \left( \frac12 (h_{11} - h_{22}) \sin (2\phi) - h_{12} \cos(2\phi) \right) \cos \theta, \\
  \label{eq:loc:1}
  & \tan{\theta} = \frac{h_{11}+h_{22}}{h_{13} \cos \phi + h_{23} \sin \phi}, \\
  \label{eq:loc:2}
  & \tan{2\phi} = \frac{2 [ h_{12} (h_{11} + h_{22}) + h_{13} h_{23} ]}{h_{11}^2 - h_{22}^2 + h_{13}^2 - h_{23}^2}.
\end{align}
Given the five measured values of $h_{ij}$ from SOGRO channels, the four unknowns such as $\theta$, $\phi$, $h_+$ and $h_\times$ can be found by solving the four equations above. Note, however, that these four equations are invariant under $\theta \to \pi - \theta$, $\phi \to \phi \pm \pi$ and $h_\times \to -h_\times$. Hence, we cannot distinguish $\hat{n}$ from $-\hat{n}$ in the solution, which is originated from the symmetry of a single SOGRO. To break this degeneracy, we need at least two detectors at separate locations.

By solving Eq.~\eqref{eq:loc:2}, we find four degenerate solutions $\phi = \phi_p + n \pi / 2$ in the range of $-\pi < \phi \le \pi$, where $\phi_p$ is a particular solution. We then obtain a solution of $\theta$ for each $\phi$ from Eq.~\eqref{eq:loc:1}. Now, $h_+$ and $h_\times$ are obtained from Eqs.~\eqref{eq:loc:hp} and \eqref{eq:loc:hc}, respectively. Note that, among the four solution pairs, only two satisfy $h_{ij} = \sum_A e^A_{ij} h_A$. The directions given by these two solutions are exactly opposite to each other, $\hat{n}_1 = -\hat{n}_2$; i.e. one is the direction to the source and the other to the GW propagation. Note also that five channels, instead of four, are needed to determine the four unknowns, which was already pointed out in the case of spherical resonant-mass GW detectors~\cite{PhysRevD.51.2517}.

\begin{figure}[tbp]
  \begin{subfigure}{.49\textwidth}
    \centering
    \includegraphics[width=.95\linewidth]{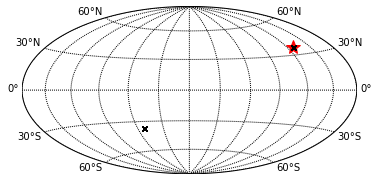}
    \caption{$h_{ij}$ only}
    \label{fig:loc:0}
  \end{subfigure}
  \begin{subfigure}{.49\textwidth}
    \centering
    \includegraphics[width=.95\linewidth]{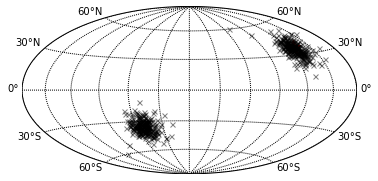}
    \caption{SNR: 167.9}
    \label{fig:loc:0.1}
  \end{subfigure}
  \newline
  \begin{subfigure}{.49\textwidth}
    \centering
    \includegraphics[width=.95\linewidth]{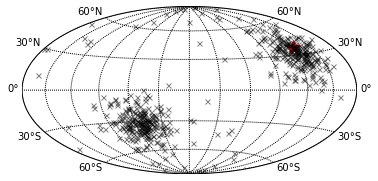}
    \caption{SNR: 83.94}
    \label{fig:loc:0.2}
  \end{subfigure}
  \begin{subfigure}{.49\textwidth}
    \centering
    \includegraphics[width=.95\linewidth]{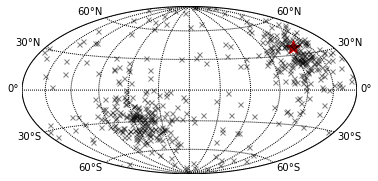}
    \caption{SNR: 55.96}
    \label{fig:loc:0.3}
  \end{subfigure}
  \newline
  \begin{subfigure}{.49\textwidth}
    \centering
    \includegraphics[width=.95\linewidth]{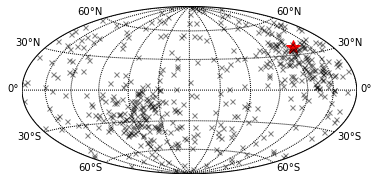}
    \caption{SNR: 41.97}
    \label{fig:loc:0.4}
  \end{subfigure}
  \begin{subfigure}{.49\textwidth}
    \centering
    \includegraphics[width=.95\linewidth]{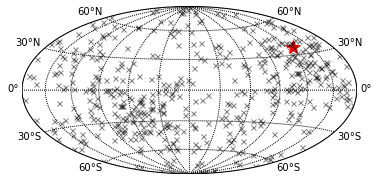}
    \caption{SNR: 33.57}
    \label{fig:loc:0.5}
  \end{subfigure}
    \caption{Localization for a simulated sine-wave signal over Gaussian noise background. We assume all five channels are working at the design sensitivity. Standard deviations of the noise are assumed to be zero (in panel a), 1/10 of the amplitude of the signal (SNR 167.9, panel b), 1/5 of the amplitude of the signal (SNR 38.94, panel c), 3/10 of the amplitude of the signal (SNR 55.96, panel d), 2/5 of the amplitude of the signal (SNR 41.97, panel e), and 1/2 of the amplitude of the signal (SNR 33.57, panel f). 
    The position of the synthetic source signal is marked with a red star. The calculated positions at every $t$ are shown as `$\times$'. The degeneracy of two {\it positions} in all panels is expected due to the symmetry of a single SOGRO response function.}
    \label{fig:localization}
\end{figure}

Next, considering noise, a pair of direction vectors, $\hat{n}(t_0)$ and $-\hat{n}(t_0)$, can be obtained at $t=t_0$ by solving Eqs.~\eqref{eq:loc:hp}-\eqref{eq:loc:2} with $h_{ij}$ replaced by $s_{ij} = h_{ij} + n_{ij}$.
The noise background is expected to shift the solution vectors from the direction of the source, but we can take the average to find the correct direction.
As a simple realization, we consider a sine wave signal, $h_+ = a_+ \cos (\omega_0 t)$ and $h_\times = a_\times \sin (\omega_0 t)$, over Gaussian noise background of distribution $\mathcal{N}(0,\sigma^2)$.
The position $(\theta_0, \phi_0)$ of the simulated signal is marked by a star in Fig.~\ref{fig:localization}.
The estimated positions are plotted as `$\times$' at every time step.
The amplitudes are set $a_+ = a_\times = 1$ and the standard deviation of the noise background is $\sigma = 0$ for Fig.~\ref{fig:localization}(a), $\sigma = 0.1$ (SNR 167.9) for Fig.~\ref{fig:localization}(b), $\sigma = 0.2$ (SNR 83.94) for Fig.~\ref{fig:localization}(c), $\sigma = 0.3$ (SNR 55.96) for Fig.~\ref{fig:localization}(d), $\sigma = 0.4$ (SNR 41.97) for Fig.~\ref{fig:localization}(e), and $\sigma = 0.5$ (SNR 33.57) for Fig.~\ref{fig:localization}(f).


\subsection{Correlation for multi detectors with multi-channels}

\noindent
In this section, we consider the data analysis required for the detection of a stochastic gravitational wave background (SGWB) using SOGRO detectors. In particular, the dimensionless spectral energy density $\Omega_{\rm gw}(f) = (d\rho_{\rm gw}/d \ln{f})/\rho_{\rm cr}$ has been analyzed, taking into account the multi-channel nature of SOGRO. Recall that the minimum detectable value of $\Omega_{\rm gw} (f)$ in a single interferometer is given by~\cite{Allen:1997ad}
\begin{equation}
  \label{eq:min:omega:component}
  \left[ \Omega_{\rm gw} (f) \right]_{\rm min} = \frac{4 \pi}{3 H_0^2} f^3 S_n (f) \frac{\rho^2}{F},
\end{equation}
where the angular efficiency factor is given by $F = \langle F_+^2 \rangle + \langle F_\times^2 \rangle = 2/5$ for terrestrial interferometers. $H_0 = h_0 \times 100\,{\rm km}/{\rm s}/{\rm Mpc}$ is the Hubble constant today. Since the off-diagonal channels in a single SOGRO detector have the same angular efficiency factor, each detection channel gives the same minimum detectable value for $\Omega_{\rm gw} (f)$. One might think of a SOGRO as multiple interferometers at the same location, from which we may have some gain. However, the overlap reduction functions between any two of the off-diagonal channels defined by
\begin{equation}
  \label{eq:overlap}
  \Gamma (f) = \int \frac{d^2\hat{n}}{4 \pi} \int \frac{d\psi}{2\pi} \left[ \sum_A F_A^{(1)} (\hat{n}) F_A^{(2)} (\hat{n}) \right] \exp \left( 2 \pi i f \hat{n} \cdot \Delta \vec{x} / c \right)
\end{equation}
are vanishing, where $\Delta \vec{x} = \vec{x}_{(2)} - \vec{x}_{(1)} = 0$ for this case.
As for the diagonal channels, on the other hand, the angular efficiency factors are given by $F = 8/15$ and the overlap reduction functions between two of three diagonal channels are calculated as $\Gamma = -4/15$.
Since we have two independent pairs of the diagonal channels, the minimum value of $\Omega_{\rm gw} (f)$ in a single SOGRO is given by
\begin{align}
  \left[ \Omega_{\rm gw} (f) \right]_{\rm min} &\sim \frac{5 \pi^2}{H_0^2} \frac{f^3 \rho^2}{(2 T \Delta f)^{1/2}} \left[ S_{n,11} (f) S_{n,22} (f) \right]^{1/2} \\
    &\simeq 8.78 \times 10^{-3} \left( \frac{0.73}{h_0} \right)^2 \left( \frac{f}{1\,{\rm Hz}} \right)^3 \left( \frac{\rho}{5} \right)^2 \nonumber \\
    & \qquad \qquad \qquad \times \left( \frac{1\,{\rm yr}}{T} \frac{10\,{\rm Hz}}{\Delta f} \right)^{1/2} \left( \frac{S_{n,{\rm diag}}}{1 \times 10^{-36} \, {\rm Hz}^{-1}} \right),
\end{align}
where $T$ is the integration time and $\Delta f$ the frequency band of the detector, assuming that the sensitivity is flat in the band. The NSD of diagonal channels are assumed to be the same, $S_{n,ii} = S_{n,{\rm diag}}$ with $i=1,2$. In this derivation, noises of different channels are assumed to be independent, {\it i.e.,} uncorrelated. The Brownian motion for each channel comes from thermal noise in different combinations of modes and the SQUID noise comes from different SQUIDs. Hence, they are generally uncorrelated. The platform noise may be correlated between some channels. The Newtonian noise coming from seismic or atmospheric density fluctuations surround may also give correlation. These correlated noises will not be suppressed by the time integration and so the upperbound of $\Omega_{\rm gw} (f)$ measured by a single SOGRO should be limited by these noises.

We now consider a network of two identical SOGRO detectors which are located at some distance $d$ from each other, but in the same orientation for convenience. We assume that their separation is far enough so that the noises of two detectors are uncorrelated, but still close enough so that the exponential factor in the overlap reduction function in Eq.~\eqref{eq:overlap} approximates to 1. Then, the overlap reduction functions between two detectors simply reduce to $8/15$ for the diagonal channels in the same direction, $-4/15$ for the diagonal channels in the different direction, and $2/5$ for the off-diagonal channels in the same direction, respectively. The NSD of the $x$-$y$ channel is roughly half of the NSDs in other channels, i.e. $S_{n,ii} \approx S_{n,i3} \approx 2 S_{n,12}$ with $i=1,2$ in SOGRO.
For two SOGRO detectors, finally, the minimum detectable value of $\Omega_{\rm gw}(f)$ approximates to
\begin{align}
  \left[ \Omega_{\rm gw} (f) \right]_{\rm min} &\sim \frac{4 \pi^2}{3 H_0^2} \frac{f^3 \rho^2}{(2 T \Delta f)^{1/2}} \left[ \frac{32}{45} S_{n,\textrm{\scriptsize diag}}^{-2} + \frac{4}{25} \left( S_{n,12}^{-2} + S_{n,23}^{-2} + S_{n,31}^{-2} \right) \right]^{-1/2}\label{eq:Omega_gw_of_stochastic_sensitivy} \\
    &\simeq 1.81 \times 10^{-3} \left( \frac{0.73}{h_0} \right)^2 \left( \frac{f}{1\,{\rm Hz}} \right)^3 \left( \frac{\rho}{5} \right)^2 \nonumber \\
    & \qquad \qquad \qquad \times \left( \frac{1\,{\rm yr}}{T} \frac{10\,{\rm Hz}}{\Delta f} \right)^{1/2} \left( \frac{S_{n,\textrm{\scriptsize diag}}}{1 \times 10^{-36} \, {\rm Hz}^{-1}} \right)~.
\end{align}
As mentioned above, however, the correlation between noises of different channels at the same detector may not be removed completely. Hence, the network of different channels at the same detector has not been taken into account in this formula.

For the SOGRO network of $N$ identical detectors, with the same orientation and the same assumption that they are still close enough to keep the overlap reduction functions unchanged, the minimum detectable value of $\Omega_{\rm gw}(f)$ 
becomes $\sqrt{2/N(N-1)}$ times Eq.~(\ref{eq:Omega_gw_of_stochastic_sensitivy}).
It should be noted here that $N$ is restricted not to be large number because of the strong assumptions --- the same orientation and close locations.


\section{Target science of SOGRO}
\label{sources}


\subsection{Coalescence of Binary Black Holes}
\label{binary}

Since the first detection of GW150914 by aLIGO\cite{Abbott:GW150914} in 2015, about 90 compact binary coalescences have been observed \cite{GWTC1,GWTC2,GWTC2-1,GWTC3}. They include stellar-mass BH-BH, NS-NS, and NS-BH binaries. The detectable mass range for LIGO-Virgo-KAGRA (LVK) detectors is less than a few hundred solar masses because their most sensitive frequency band is about $20-2000$ Hz. 

More massive binary black holes (BBHs) are the prime target for SOGRO, given its coverage of the frequency band between 0.1 and 10 Hz. The highest GW frequency in the inspiral phase---innermost stable circular orbit (ISCO)---is inversely proportional to the binary's total mass as follows\cite{Abadie:2010},
\begin{equation} \label{eq:f_isco}
f_{GW,ISCO}=\frac{1}{\pi}\bigg(\frac{1}{6}\bigg)^{1.5}\frac{c^3}{GM} \sim \frac{4396}{M/M_{\odot}} ~[\rm {Hz}]~.
\end{equation}
Hence, SOGRO can observe only the inspiral phase of compact binaries with masses less than approximately 440 M$_{\odot}$. Conversely, for binaries with masses exceeding 44,000 M$_{\odot}$, only the merger and ringdown phases are observable. Even when considering the merger and ringdown phases, the observable mass range with SOGRO is limited up to roughly $10^5$ M$_{\odot}$.

\begin{figure}
    \centering    \includegraphics[width=15cm,height=12cm]{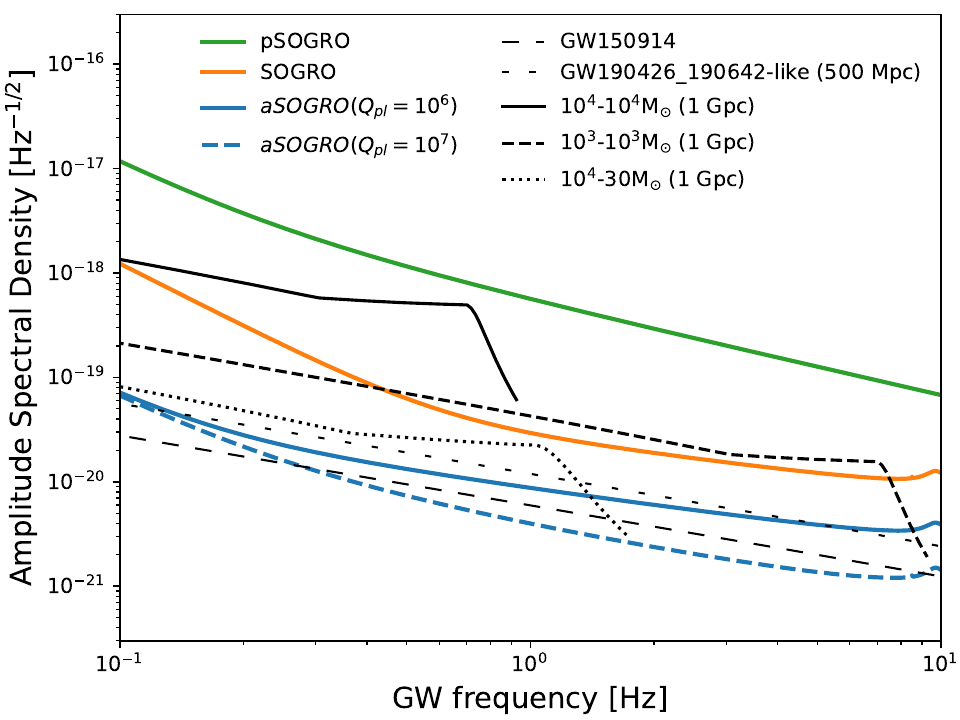}
    \caption{Amplitude spectral densities (ASDs) of pSOGRO, SOGRO, and aSOGRO are overlaid with the expected signals from BBHs with different masses. The inclinations of BBHs are assumed to be zero (face-on). GW150914 signal is based on its observed masses and distance given in the literature\cite{GWTC1}. GW190426\_190642-like source is a BBH whose masses are adopted from GW190426\_190642 (105.5 and 76.0 M$_{\odot}$) but located at 500 Mpc which is closer than its actual distance (4.58 Gpc)\cite{GWTC2-1}. For the BBHs with IMBH, the masses and distances are assumed to be as indicated in the figure.}
    \label{fig:asd}
\end{figure}

BHs with masses in the range $\sim 10^{2}-10^{5}$ M$_{\odot}$ are termed intermediate-mass black holes (IMBHs). The mass range of the targeted compact binaries of SOGRO overlaps with that of the IMBHs. The existence of IMBHs can be inferred from the straightforward extrapolation of the well-known correlation between a galaxy bulge's stellar velocity dispersion and the mass of its supermassive BH \cite{Merritt:2000, Ferrarese:2000, Gebhardt:2000, McConnell:2011, Kormendy:2013}. Although several observations support the existence of IMBHs \cite{Patruno:2006, Maccarone:2007, Chilingarian:2018, Takekawa:2019, Woo:2019}, their existence still remains a topic of debate \cite{greene:2020}.

\begin{figure}
    \centering
    \includegraphics[width=15cm,height=12cm]{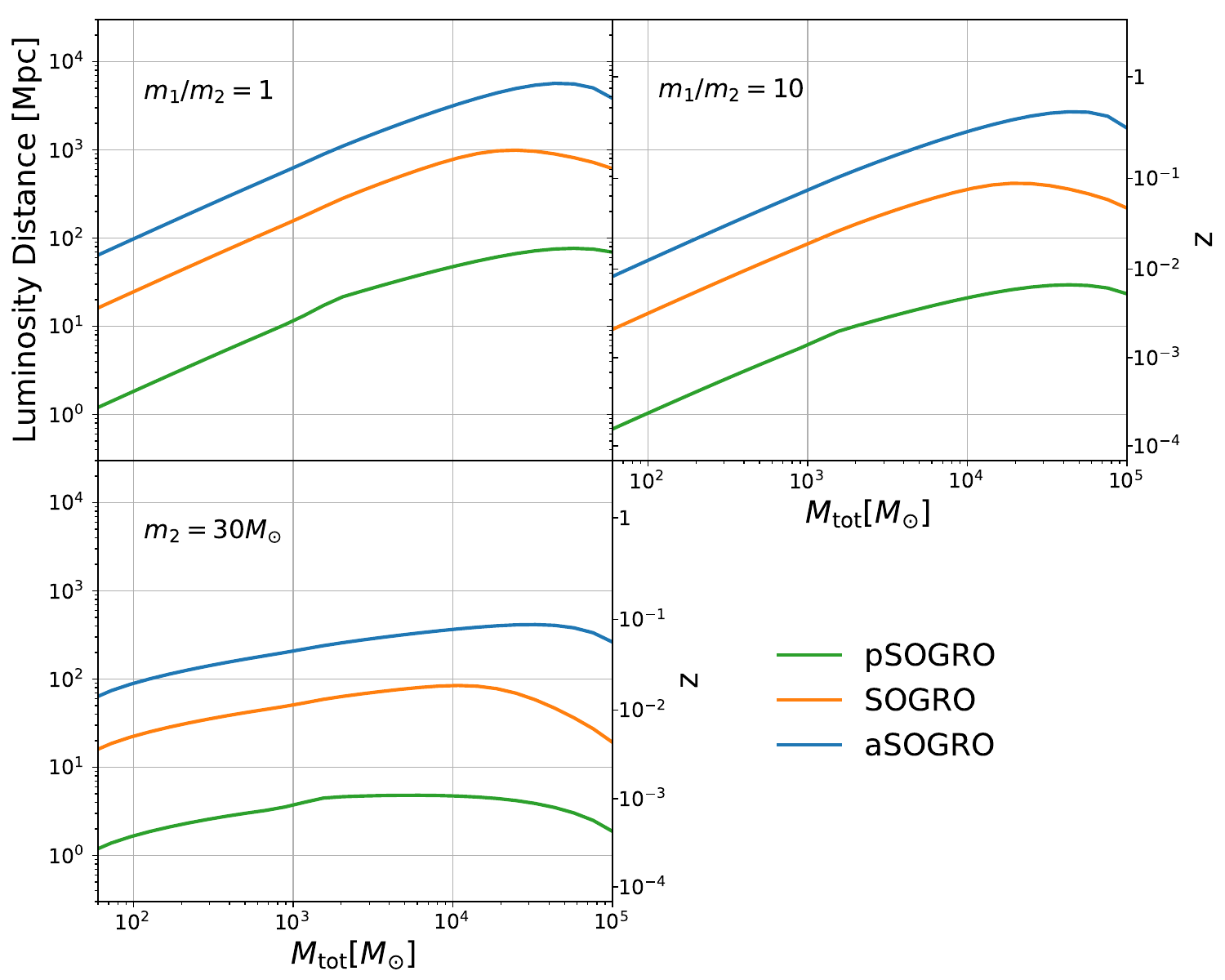}
    \caption{Horizon distances versus the total mass of BBHs based on the sensitivities of pSOGRO (green), SOGRO (orange), and aSOGRO (blue), respectively. We consider the cases where the mass ratios are $m_{1}/m_{2} = 1$ and 10, and additionally when the secondary mass is fixed to be 30M$_{\odot}$. All BBHs are assumed to be face-on and $\rho=8$.}
    \label{fig:horizon}
\end{figure}

In this section, we evaluate the detectability of IMBH binaries (IMBHBs) by considering SOGRO's sensitivity and estimate their detection rates based on an assumption of their formation in star clusters. We utilize analytical waveforms encompassing inspiral, merger, and ringdown phases with non-precessing BH spins \cite{Ajith:2011}. Furthermore, we adopted the cosmological parameters from the Planck mission \cite{Planck:2016} to compute the distance scale.

Fig.~\ref{fig:asd} displays the sensitivity curves of pSOGRO, SOGRO, and aSOGRO obtained from Table.~\ref{tab:DP} and Eq.~(\ref{nsd:optimal}), represented by green, orange, and blue solid lines, respectively.\footnote{It should be pointed out that the definition of the detector sensitivity in Eq.~\eqref{nsd:optimal} differs from that in Ref.~\cite{Paik2020IJMPD}, which is simply the sum of all five channels. The definition in Ref.~\cite{Paik2020IJMPD} is neither the optimal NSD nor the direction averaged one. The sensitivity curve for the aSOGRO with the platform quality factor of $Q_{\rm pl}=10^7$ is also plotted in blue dashed line for comparison.}
Here the plot is the sum of the antenna thermal, amplifier and platform thermal noises, and the noise models introduced in Ref.~\cite{Paik2020IJMPD} are used with the design parameters in Table~\ref{tab:DP}. The seismic and Newtonian noises are assumed to be suitably mitigated below the sensitivity curves. For the pSOGRO, its platform thermal strain noise is newly obtained as described in Sect.~\ref{platform} below.
Black lines denote the amplitude spectral densities (ASDs) for BBHs with various masses and mass ratios. Given the observed samples, aSOGRO's sensitivity can reach the ASDs of stellar mass BBHs with masses similar to those of GW150914 or GW190426\_190642 within its detection frequency band, with a SNR of $\rho\sim (1-2)$. Here, we assume a GW190426\_190642-like BBH has the same masses as GW190426\_190642, which is the most massive BBH in GWTC catalogs, but is located at 500 Mpc. For IMBHs, we simplify the source models by assuming all BBHs are at 1 Gpc. Only aSOGRO has the capability to detect all IMBHBs considered in Fig.~\ref{fig:asd}. The sensitivities of pSOGRO and SOGRO are included as references for comparison with aSOGRO. SOGRO can also detect a $10^{4}-10^{4}$ M$_{\odot}$ binary at 1 Gpc with $\rho\simeq 8$.

Fig.~\ref{fig:horizon} illustrates the horizon distances of pSOGRO, SOGRO, and aSOGRO as functions of the total mass of a BBH. Here, the horizon distance is defined as the maximum observable distance when the BBH is located at the zenith of the SOGRO with zero inclination for a given SNR $\rho=8$. The overall shape of the horizon distance is determined by the sensitivity curve's shape and the frequency range of GW signals from the BBHs. Since massive BBHs merge before reaching 10 Hz, the horizon distance decreases in the regime of the large total mass. The optimal total mass for detecting BBHs, corresponding to the expected SOGRO sensitivities is roughly a few tens of thousands of solar masses. The maximum horizon distances of SOGRO and aSOGRO for equal mass BBHs are approximately $z\simeq 0.2$ and 0.87 respectively, corresponding to the luminosity distance of roughly 1 Gpc and 6 Gpc. These values match or surpass the distances of BBHs observed by current interferometric GW detectors. We also calculated the horizon distance of IMBH-IMBH binaries with a mass ratio of 10, and binaries of IMBH with a 30 M$_{\odot}$ stellar mass BH (sBH), which yield smaller horizon distances than equal mass IMBH-IMBH binaries due to their smaller chirp masses. The horizon distance of aSOGRO for the IMBH-sBH extends up to $z\simeq 0.1$. It should be noted that the mass ratio in the IMBH-sBH plot is dependent on the horizontal axis.

To estimate the detection rate of IMBHBs, we must first make assumptions regarding their formation history. While several formation mechanisms have been proposed \cite{Giersz:2016}, our understanding of them remains limited. In this study, we adopt the methodology of earlier research \cite{Fregeau:2006, Amaro-Seoane:2010}. These studies postulate that IMBHs originate from the collapse of a very massive star, which itself forms via a runaway collision of massive main-sequence stars in a young dense star cluster \cite{Portegies_Zwart:1999, Ebisuzaki:2001, Portegies_Zwart:2002, Portegies_Zwart:2004, Gurkan:2004}. In this cluster, two independently formed IMBHs can be gravitationally bound to become a binary, which subsequently merges in a short time. 

The overall process of estimating the detection rate is as follows. We use the star formation rate in mass ($M_{\rm SF}$) per unit comoving volume ($V_{c}$) per unit time at the detector ($t_{0}$) to find the number of star clusters in which the IMBHBs are formed among them and integrate it up to the detectable volume which is calculated by using angle averaged sensitivity (Eq.~\ref{nsd:averaged}). If we consider the distribution function of the mass of the clusters $f(M_{\rm cl})$ in calculating the number of clusters, the detection rate $R$ can be expressed as follows (Eq.(2) and Eq.(4) in Ref. \cite{Fregeau:2006})

\begin{equation}\label{eq:detection_rate}
    R = \int^{z_{\rm max}}_{0} \frac{d^2 M_{\rm SF}}{dV_{\rm c}dt_{e}} \frac{dt_{e}}{dt_{0}} g_{cl}g \int^{M_{\rm cl,max}(z)}_{M_{\rm cl,min}(z)}\frac{f(M_{\rm cl})}{\int{M_{\rm cl}f(M_{\rm cl})dM_{\rm cl}}}dM_{\rm cl} \frac{dV_{\rm c}}{dz}dz ~,
\end{equation}
where $t_{e}$ is the local time at the source, and the parameters $g_{\rm cl}$ and $g$ are the fraction of the mass included in the star cluster and the fraction of the clusters that can produce IMBHBs, respectively.

We use three different models for the star formation rates \cite{Madau:2000,Steidel:1999,Blain:1999} adopted in Ref. \cite{Porciani:2001}. But these models do not give much difference in detection rates (about 10\% at most), because their differences are conspicuous out of the horizon distance of the aSOGRO. Thus we use the average value of them. For the fraction parameters that are the most uncertain factors, we consider the conservative and optimistic ranges from the literature as $g_{\rm cl}\sim(0.0025-0.1)$ \cite{McLaughlin:1999,Fregeau:2006,Amaro-Seoane:2010} and $g\sim(0.1-0.5)$ \cite{Fregeau:2006,Freitag:2006a,Freitag:2006b,Amaro-Seoane:2010}. In addition, we adopt the distribution function of the mass of the cluster as $f(M_{\rm cl}) \propto M_{\rm cl}^{-2}$ from the Ref. \cite{Zhang:1999}. We assume the lowest mass of the IMBHB in the mass distribution to be 200 M$_{\odot}$ \cite{Amaro-Seoane:2010}, and the mass ratio between the IMBHB and the star cluster to be $M_{\rm BBH}/M_{\rm cl}=2\times10^{-3}$ \cite{Gurkan:2004}, which means the lower mass limit of the cluster that can have IMBHB is $10^{5}$ M$_{\odot}$. The lower limit in the integration of the cluster mass in Eq.~\eqref{eq:detection_rate} is selected as the larger one between $10^{5}$ M$_{\odot}$ and the minimum cluster mass considering the detector's sensitivity and the distance to the source, since the detectable IMBHB's mass is larger than 200 M$_{\odot}$ when the source is far away. The upper limit of the integration is also chosen as the smaller one between $10^{7}$ M$_{\odot}$, which is the upper limit of the cluster mass considered in Ref.~\cite{Zhang:1999}, and the maximum cluster mass detectable at a given distance. Thus, the IMBHB's mass range used in the calculation of the detection rate is $2\times10^{2}-2\times10^{4}$ M$_{\odot}$. But the SOGRO can detect the BH binaries with higher mass (Fig.~\ref{fig:horizon}), the same trend of the mass distribution is extended to the larger masses when the reference and optimistic detection rates are estimated.

\begin{table}[b]
\caption{Estimated detection rates for equal-mass BBH mergers. The most conservative ($R_{\rm low}$), reference ($R_{\rm ref}$), and the most optimistic ($R_{\rm high}$) estimates are presented with the model assumptions. We present the detection rates obtained for two values of the SNR for comparison: $\rho=8$ (without parentheses) and $\rho=5$ (within the parentheses). Details of model assumptions can be found in the main text.}
\centering
\label{tab:detection_rate}
\begin{center}
\begin{tabular}{M{2cm}|M{4cm}|M{4cm}|M{4cm}}

    \hline
          &  R$_{\rm low}$  &  R$_{\rm ref}$  &  R$_{\rm high}$  \\    
    \hline\hline
    \multirow{5}{*}{}  & \multicolumn{3}{c}{$\rho=8$ (5), equal mass, no spin, randomly oriented source} \\ 
    \cline{2-4}
                       & \footnotesize{$M_{\rm BBH,max}=2\times10^{4}$ M$_{\odot}$} & \footnotesize{$M_{\rm BBH,max}=10^{5}$                           M$_{\odot}$} & \footnotesize{$M_{\rm BBH,max}=10^{5}$ M$_{\odot}$} \\
    Model              & $g_{\rm cl}=0.0025$ & $g_{\rm cl}=0.01$ & $g_{\rm cl}=0.1$ \\
    assumption         & $g=0.1$ & $g=0.2$ & $g=0.5$ \\
                       & $R_{\rm double}/R_{\rm single}=0.1$ & $R_{\rm double}/R_{\rm single}=0.5$ & $R_{\rm double}/R_{\rm single}=1$ \\
    \hline
    \small{Detection Rate (yr$^{-1}$)} & 0.0014 (0.0043) & 0.028 (0.076) & 0.95 (2.5) \\
    \hline

\end{tabular}
\end{center}
\end{table}

The above calculation assumes that the IMBHBs are formed in a single-star cluster. However, two IMBHs that are formed in different star clusters can merge by the collision of the star clusters. It is known that about ($10-100$)\% of the number of IMBHBs can be added through this double cluster scenario \cite{Amaro_Seoane:2006}. Thus the final estimation of the detection rate is presented by multiplying the above calculations by 1.1 and 2 for the conservative and the optimistic estimations, respectively. The estimations of the detection rate of SOGRO are presented in Table~\ref{tab:detection_rate}.

We have categorized the estimated detection rate of aSOGRO into three cases---the most conservative, the most optimistic, and a reference value between them. The conditions for each of these categories are detailed in Table.~\ref{tab:detection_rate}. Additionally, we considered cases with $\rho=5$, which are indicated in parentheses. In our estimations of the detection rates, we assumed the averaged combined sensitivity of aSOGRO and the random orientation of BBHs.

With aSOGRO, in the most optimistic scenario, we anticipate approximately 1 to 2.5 detections per year. However, in the least favorable scenario, the detection of IMBHs becomes nearly infeasible. The most likely detection rates lie between 0.03 to 0.08 per year. As it stands, the sensitivity of aSOGRO may need further enhancement for IMBHB detections. However, it's still premature to make a definitive conclusion due to the significant gap between the conservative and the optimistic estimates, stemming from the limited information on IMBHs.

We can also estimate the detection rate of IMBH-sBH binaries since their GW frequency is dependent on their total mass (see Eq.~\ref{eq:f_isco}). However, their horizon distances are significantly shorter than those of IMBH-IMBH binaries because their chirp masses are dominated by the sBH. Considering that the observable volume is proportional to the cube of the horizon distance, this results in a significantly reduced detection rate. Moreover, considering that only about 1\% \cite{sedda:2020} of IMBHs are paired with a sBH, the rate would be even lower. Although the formation scenario wherein IMBHs grow via successive mergers with sBHs could potentially enhance the detection rate \cite{gerosa:2017,Shinkai:2017}, it remains doubtful that it would lead to a substantial increase in SOGRO's detection rates, because the mass of BBH in the early growth stages is small, implying a limited observable volume.

The SOGRO's frequency band encompasses the inspiral phase of binaries made up of two sBHs or NSs, even though their merger and ringdown phases fall outside of this range. The inspiral signals from these sources linger within the SOGRO's frequency band for an extended period. For instance, a 30-30 $M_{\odot}$ BBH remains in the $0.1-10$ Hz range for approximately two weeks, which is equivalent to roughly $10^5$ cycles \cite{nakamura:2016}. Detecting such signals would be invaluable for accurately estimating source parameters and testing general relativity \cite{sedda:2020}. Moreover, it could be beneficial for higher frequency detectors and electromagnetic follow-ups by providing early warnings. For a randomly oriented equal-mass (30-30 M$_{\odot}$) BBH, the observable comoving volume of aSOGRO is approximately $4.5\times10^{-4}$ Gpc$^{-3}$ at $\rho=5$. Taking into account the event rates observed with LIGO and Virgo during the first half of their third observing run \cite{GWTC2}, this translates to a detection rate of 0.01 yr$^{-1}$.

\begin{figure}
    \centering    \includegraphics[width=16cm,height=6.5cm]{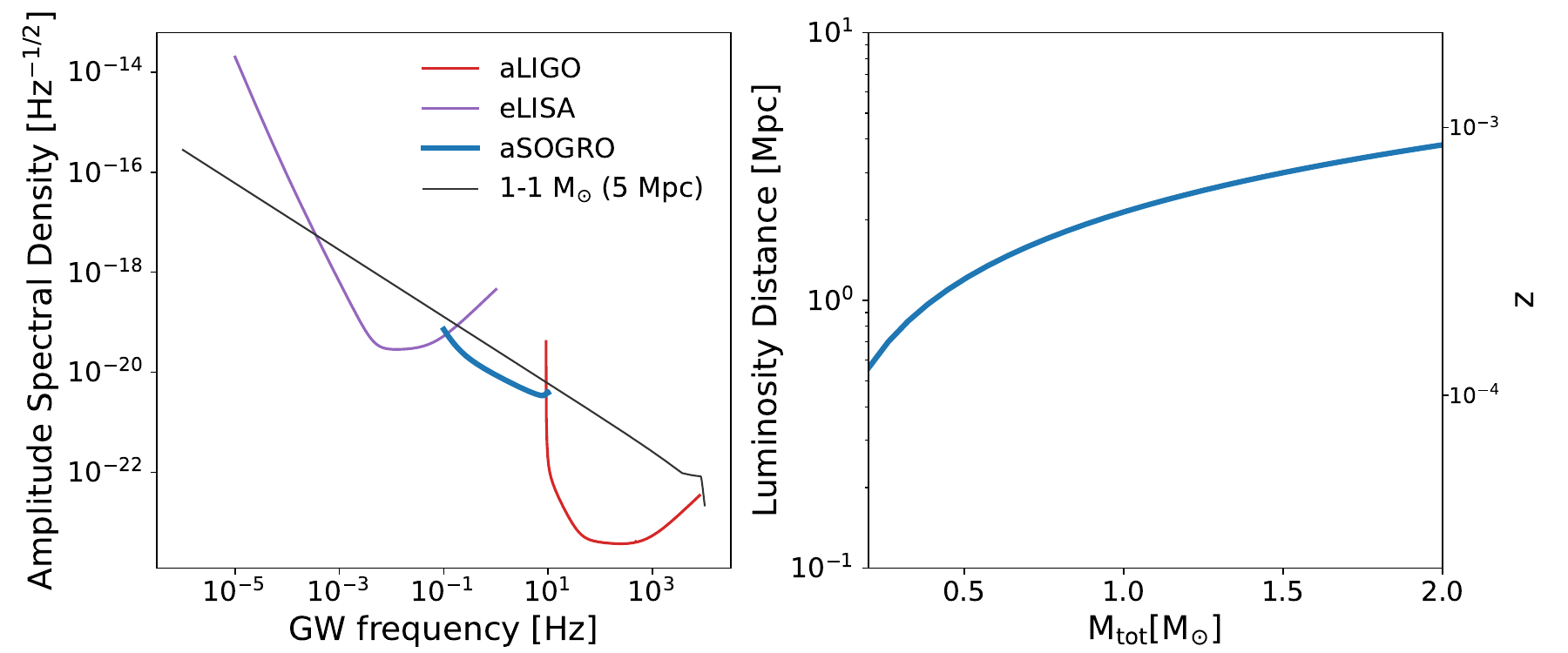}
    \caption{Left: Expected GW signal from an equal-mass  (1 M$_{\odot}$) BBH coalescence and the sensitivity curves of aLIGO, eLISA, and aSOGRO. Right: Horizon distance with respect to the total mass of BBHs when the mass ratio is one. The curve is based on the fixed SNR $\rho=8$.}
    \label{fig:subsolar}
\end{figure}

We can also explore the GWs emitted from the inspiral phase of sub-solar mass BBHs. They are not formed through stellar evolution. Some theories suggest that early universe density fluctuations might spawn primordial BHs \cite{carr:2020}, while others hint at the production by non-baryonic dark matter particles \cite{shandera:2018}. Fig.~\ref{fig:subsolar} compares the ASD for a 1-1 M$_{\odot}$ with those of detectors, and depicts the horizon distance of aSOGRO with respect to the total mass of the sub-solar mass BBH's. Due to their small chirp masses, aSOGRO's horizon distance for sub-solar BBHs is limited to several Mpc. This corresponds to the size of the Local Group which includes the Milky Way and M31. Constraints from the first half of the advanced LIGO and Virgo's third observing run imply a merging rate of about 1200 Gpc$^{-3}$ yr$^{-1}$ for 1-1 M$_{\odot}$ sub-solar mass BBHs \cite{Nitz:2021}. Taking into account this merger rate as face values, we anticipate aSOGRO's detection rate for sub-solar mass BBHs to be no more than roughly $10^{-4}$ yr$^{-1}$. 

\begin{figure}
\begin{center}
\includegraphics[width=0.9\textwidth]{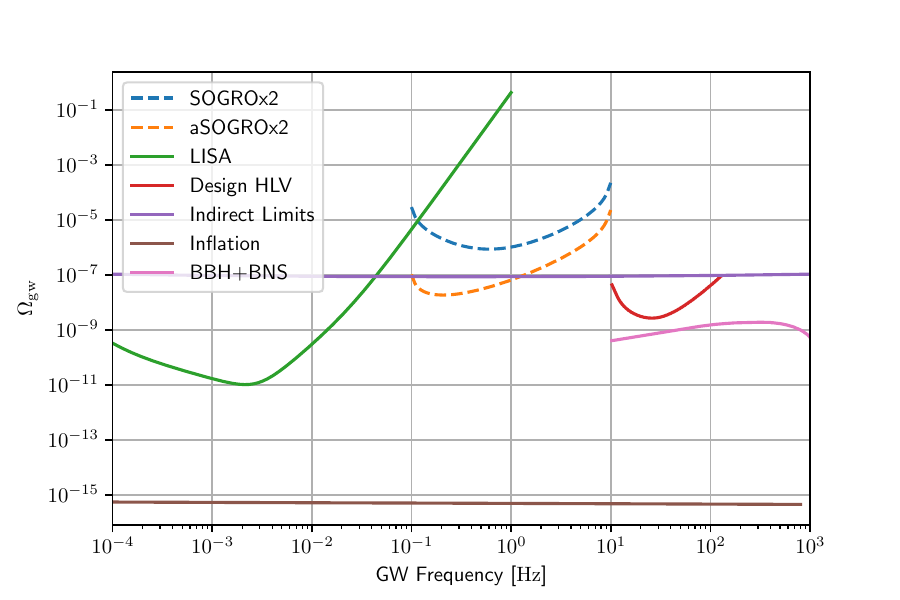}
\end{center}
\caption{Power-law integrated (PI) sensitivity curves for different detectors for SGWB. In addition, theoretical predictions and constraints are shown.  The two-detector configurations with SOGRO (blue dotted line) and aSOGRO (orange dotted line) are presented in a frequency range of $0.1\,\mathrm{Hz}\leq f\leq10\,\mathrm{Hz}$.}
\label{fig:SGWB_sensitivity}
\end{figure}


\subsection{Stochastic background}
\label{stochastic}

In the two-detector correlation as in Eq.~(\ref{eq:Omega_gw_of_stochastic_sensitivy}), let us assume power-law model for SGWB $\Omega_{\mathrm{gw}}\left(f\right)=\Omega_{0}\left(f/f_{0}\right)^{\alpha}$ for real $\alpha$. Then, the minimum detectable model with SNR $\rho$ is given by
\begin{equation}
    \Omega_{\mathrm{gw}}^{\mathrm{min}}\left(f\right)=\frac{4\pi^{2}}{3H_{0}^{2}}\frac{\rho^{2}}{\sqrt{2T}}\left[\int_{0}^{\infty}df'f'^{2\left(\alpha-3\right)}S_{n,\mathrm{cor}}^{-2}\left(f'\right)\right]^{-1/2}f^{\alpha},
\end{equation}
where
\begin{equation}
    S_{n,\mathrm{cor}}\left(f\right)=\left[\frac{32}{45} S_{n,\mathrm{diag}}^{-2}\left(f\right)+\frac{4}{25}\left(S_{n,12}^{-2}\left(f\right)+S_{n,23}^{-2}\left(f\right)+S_{n,31}^{-2}\left(f\right)\right)\right]^{-1/2}.
\end{equation}
Here, all detections by pairs of channels at the same detector are ignored due to the potential noise correlation among nearby channels. Overlapping the curves of minimum detectable models over a certain range of $\alpha$, we obtain the power-law integrated (PI) sensitivity curve \cite{thrane_sensitivity_2013}.

Fig.~\ref{fig:SGWB_sensitivity} shows PI sensitivity curves for several SOGRO types in a frequency range of $0.1\,\mathrm{Hz}\leq f\leq10\,\mathrm{Hz}$. As possible candidates of SGWB near the SOGRO's frequency band, we depict rough levels of a cosmological SGWB generated during the inflation era from \cite{gwplotter} and an astrophysical SGWB \cite{PhysRevD.104.022004} by generated BBH and binary neutron stars (BNS). We also plot the indirect limits \cite{PhysRevX.6.011035}, which is the upper bound of SGWB reconciled with the concordance cosmological model. For comparison, we show the sensitivity of LISA \cite{Cornish_2017} in $10^{-4}\,\mathrm{Hz}\leq f\leq1\,\mathrm{Hz}$  and the PI sensitivity for the 3-detector network consisting of two aLIGOs and Virgo \cite{PhysRevD.104.022004}, labeled as Design HLV, in $10\,\mathrm{Hz}\leq f\leq130\,\mathrm{Hz}$. For all PI curves, we use the power range $-10\leq\alpha\leq10$, the observation time interval $T=2\,\mathrm{yr}$, the SNR $\rho=\sqrt{2}$, and the Hubble constant $H_{0}=68\,\mathrm{km}/\mathrm{s}/\mathrm{Mpc}$. The curve labeled as `Inflation' is the strength of SGWB from cosmological inflation \cite{gwplotter} with the tensor-to-scalar ratio $r=0.1$ \cite{planck2018}. The curve labeled as `BBH+BNS' is the strength of astrophysical SGWB from binary black holes and binary neutron stars \cite{PhysRevD.104.022004}. The curve labeled as `Indirect Limits' shows the upper bound of SGWB constrained by cosmological observations \cite{PhysRevX.6.011035}. The SOGRO frequency band is complementary with other detectors and/or constraints. We expect that LISA, aLIGO, and aSOGRO can explore SGWB beyond the cosmological indirect limits in their own frequency range.

\begin{figure}[ht]
\begin{center}
\includegraphics[width=0.9\textwidth]{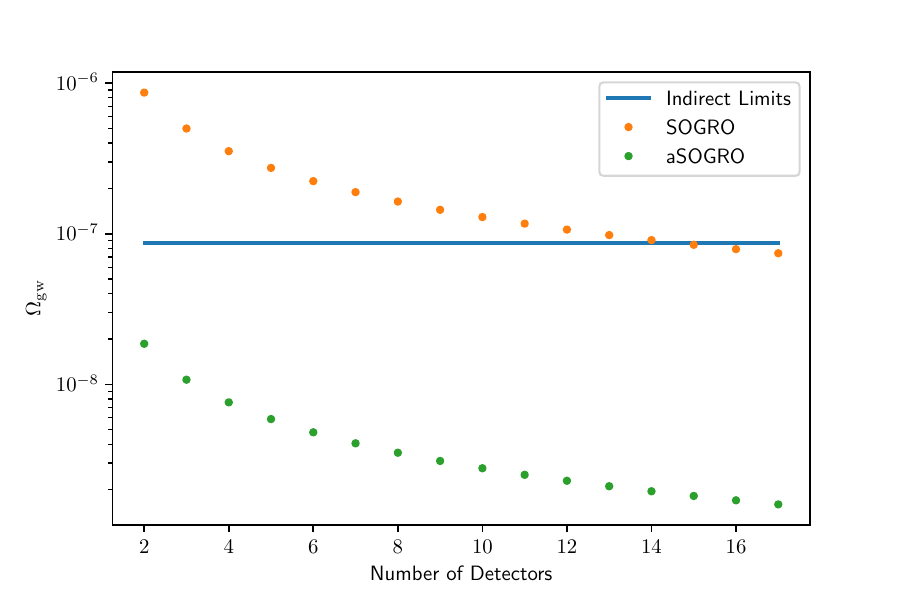}
\end{center}
\caption{We compare expected sensitivities for SGWB observation considering network of N detectors. We present results for networks consisting of N SOGRO's or N aSOGRO's. We assume that SGWB follows the power-law model with $\alpha=0$.}
\label{fig:correlation_sensitivity}
\end{figure}

If we have multiple detectors of number $N$, the observational time $T$ can be effectively increased to $T\times N\left(N-1\right)/2$ by correlating all pairs of data from $N$ detectors. Fig.~\ref{fig:correlation_sensitivity} shows sensitivity behaviors of multiple SOGRO and aSOGRO for the power-law model parameter $\alpha=0$. Orange and green dots represent the networks consisting of $N$ SOGRO and $N$ aSOGRO, respectively. The horizontal solid line shows a level of indirect limits given in Fig.~\ref{fig:SGWB_sensitivity} for the power-law model parameter $\alpha=0$. Although roughly 67 SOGRO detectors are required for the equivalent performance of two aSOGRO network, a relatively small number (roughly 15) of SOGRO detectors can touch the indirect limit.


\section{Conceptual design for a prototype SOGRO}
\label{conceptualdesign}

\noindent
In this section, we propose a prototype SOGRO, to be labeled as pSOGRO hereafter and describe its conceptual design and properties. The SOGRO concepts considered in this work have large-scale platforms with beam lengths of 30 m, 50 m, or 100 m. In addition, all these are required to be built in a deep underground site to reduce Newtonian noises, which indicates lots of technical challenges with expensive civil engineering. Therefore, it would be necessary to have a rather small-size SOGRO with which one can test plausibility and various technologies that are essential in SOGRO detectors. We summarize design parameters, platform structural analyses, and cooling systems for the pSOGRO.

\begin{table}[ht]
\caption{Design parameters for various detector concepts. Note that, for the aSOGRO, Paik {\it et. al.} have considered the arm length of 100-m as well in Ref.~\cite{Paik2020IJMPD}. }
\label{tab:DP}
\centering 
\small
\begin{tabular}{|M{3cm}|M{2cm}|M{1.9cm}|M{1.9cm}|M{1.9cm}|M{3.2cm}|}
\hline
{\bf Parameter} & {\bf SGG} & {\bf pSOGRO} & {\bf SOGRO} & {\bf aSOGRO} & {\bf Main feature} \\
\hline\hline
Individual \newline test mass $M$ (kg) & $0.10$ & $100$ & $5000$ & $5000$ & Multiple-layer Nb shell \\
\hline
Arm length $L$ (m) & $0.135$ & $2$ & $50$ & $50$ & Rigid platform \\
\hline
Antenna temperature $T$ (K) & $4.2$ & $0.1$ & $4.2$ & $0.1$ & ${\rm LHe}/{\rm He}^3-{\rm He}^4$ dilution refrigerator \\
\hline
Platform temperature $T_{\rm pl}$ (K) & $4.2$ & $0.1$ & $4.2$ & $4.2$ & Large cryogenic chamber and cooling system \\
\hline
Platform quality factor $Q_{\rm pl}$ &  & $10^6$ & $10^5$ & $10^6$ & Al platform structure \\
\hline
DM frequency $f_{\rm D}$ (Hz) & $0.02$ & $0.01$ & $0.01$ & $0.01$ & Magnetic levitation (horizontal only) \\
\hline
DM quality factor $Q_{\rm D}$ & $2 \times 10^6$ & $10^8$ & $10^7$ & $10^8$ & Surface polished pure Nb \\
\hline
Pump frequency $f_{\rm p}$ (kHz) &  & $50$ & $50$ & $50$ & Tuned capacitor bridge transducer \\
\hline
Amplifier noise no. $n$ &  & $5$ & $20$ & $5$ & Two-stage dc SQUID cooled to $0.1$ K \\
\hline
Detector noise $S^{1/2}_{\rm h}(f)$ (${\rm Hz}^{-1/2}$) & $1.0 \times 10^{-12}$ & $5.7\times 10^{-19}$ & $2.9\times 10^{-20}$ & $8.7\times 10^{-21}$ & Evaluated at $1 {\rm Hz}$ \\
\hline
\end{tabular}
\end{table}


\subsection{Design parameters}
\label{design}

The pSOGRO detector mainly consists of a platform, test masses, SQUID sensors, and a cooling system as in the SOGRO and aSOGRO. Table~\ref{tab:DP} shows the main design parameters proposed for the prototype SOGRO, and those for the SOGRO and advanced SOGRO~\cite{Paik:2016aia,Paik2020IJMPD} are also given for comparisons. The first column shows the design parameters applied to the superconducting gravity gradiometer (SGG) being developed at the University of Maryland by presumably using current technologies available~\cite{2002RScI...73.3957M}. The guiding principle in designing the pSOGRO is to extract the limits and potential problems of the SGG technologies at laboratory tests and develop core technologies for the full-scale SOGRO concepts. The proposed 2 m arm length and $100$ kg for each test mass of the pSOGRO are about $15$ and $1000$ times larger than those of the SGG, respectively.

The purpose of proposing this down-sized detector is to develop core technologies for the full-scale SOGRO and to demonstrate its feasibility. As can be seen on Table~\ref{tab:DP}, the pSOGRO has arm lengths and test masses much smaller than those of the aSOGRO, and can be developed in a common laboratory. Notice that, although this 2 m SOGRO is much smaller than the full 50 m SOGRO, implementing it would be a big challenge compared to the 19 cm or 13 cm gravity gradiometers appeared in Refs.~\cite{2002RScI...73.3957M,PhysRevApplied.8.064024}, and has never been realized so far.
The material for the arm is Al 5083, which is the same as that of the aSOGRO. The diameter and wall thickness of the circular shape tube for the arm and strut are given in Table~\ref{platformDP}.
Other design parameters for the operation temperatures of antenna and platform, quality factors, and SQUID amplifier noise are set to be the same as those of aSOGRO because these core technology components should be tested in order for the aSOGRO to be realized and their operational conditions required should be achieved in advance. Note that, with all of such design parameters, the detector noise of pSOGRO has $S_h^{1/2} \sim 5.7\times 10^{-19} {\rm Hz}^{-1/2}$ at $1 \,{\rm Hz}$, implying less sensitivity than that of the aSOGRO by about two orders of magnitude.

\begin{table}[tb]
\caption{Platform design parameters for pSOGRO. Those of the 50 m aSOGRO are also shown for comparisons.}
\label{platformDP}
\centering 
\begin{tabular}{|M{4cm}|M{5cm}|M{5cm}|}
\hline 
{\bf Parameter} & {\bf pSOGRO} & {\bf aSOGRO}  \\
\hline\hline
Arm length  & $2$ m & $50$ m  \\
\hline
Each test mass & $100$ kg & $5$ ton  \\
\hline
Material & Al 5083 & Al 5083  \\
\hline
Arm shape and size & Circular tube with $50$ cm diameter & Circular tube with $2.5$ m diameter \\
\hline
Strut shape and size &  Circular tube with $50$ cm diameter & Circular tube with $2.5$ m diameter \\
\hline
Wall thickness & $5$ mm & $1$ cm   \\
\hline
Total mass & $873$ kg  & 219 ton (excluding test masses)\\
\hline
\end{tabular}
\end{table}


\subsection{Platform structural analysis}
\label{platform}

The platform is one of the essential parts of SOGRO detector. It supports all six test masses through magnetic levitation. Various sensors and coils controlling the positions of test masses are mounted on it. The whole platform structure will be suspended by a single string or a few number of strings in a vacuum chamber which is cooled down to extremely low temperatures, {\it e.g.}, $0.1$ K. When the relative motion between a test mass and a sensor caused by a passing GWs is to be measured, any motion of the platform could cause noise because sensors are mounted on it. Thermal fluctuations of the platform could generate motions of sensors, which in turn produce some amount of signals effectively even if test masses at the ends of platform arms do not move. Such noise is called platform thermal noise. The circuit of SOGRO sensors is designed in such a way that a common motion of two test masses does not give any signal, the so-called CM (Common Mode) rejection, whereas only relative motions between two test masses ({\it i.e.}, Differential Modes) give signals to be measured.
Accordingly, thermal fluctuations producing a common motion of the platform arms would generate little noises due to the CM rejection. Similarly, small motions of the platform as a whole body coming from the suspending string would not generate serious noises.

\begin{figure}[tb]
\begin{center}
\includegraphics[width=0.8\textwidth]{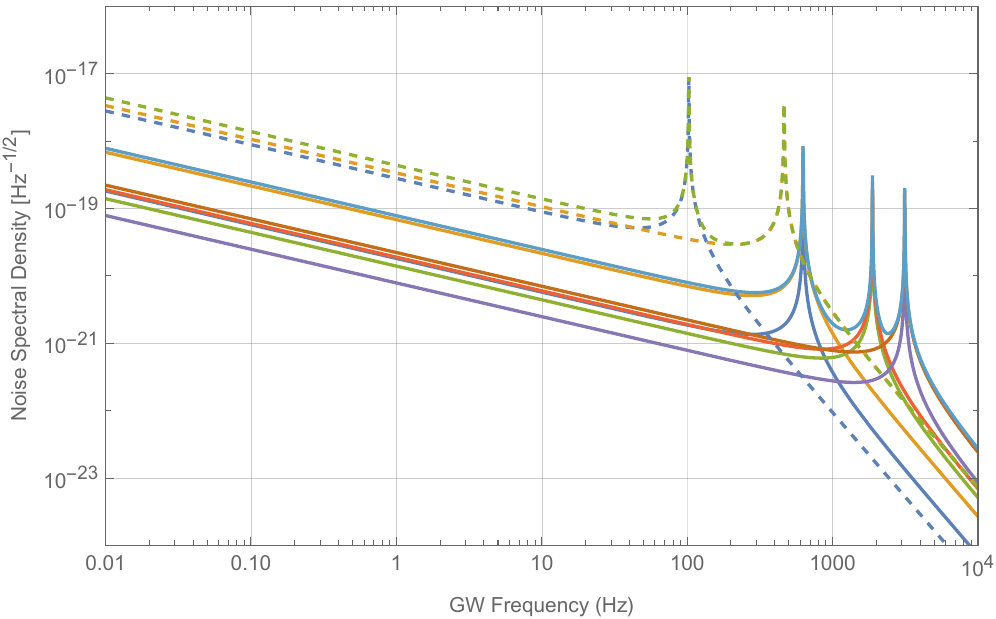}
\end{center}
\caption{Platform thermal strain noises of pSOGRO for the  first six lowest XX-modes (solid) and first two lowest XY-modes (dashed). $T_{\rm pl}=0.1$ K and $Q_{\rm pl}=10^6$. The upper-most curve is the sum.
}
\label{PTN2m}
\end{figure}

Pre-stressed modal analyses of the 2 m pSOGRO platform are performed by using Finite Element Method (FEM) with ANSYS software program. The platform structure was modeled simply using line bodies with circular tube cross-sections. Each test mass was treated as a static force, corresponding to its mass, in the negative $z$-direction acting on each single endpoint of three platform arms. Additional mass and stiffness that would be contributed by coil forms for magnetic levitations and sensors were ignored for simplicity. The whole platform structure was assumed to be suspended at the center with infinite stiffness in the $z$-direction and free motions in other degrees of freedom. Among various types of vibrations of the platform, relevant ones of interest are differential vibrations along the directions of platform arms and angular differential vibrations of two arms on an $xy$-, $yz$-, or $zx$-plane. We call those ``XX in-line modes'' and ``XY cross-component modes'' (or ``scissor modes''), respectively. The platform strain noises for such modes are shown in Fig.~\ref{PTN2m} for several resonance frequencies. One can see that the lowest resonance frequencies of platform thermal fluctuations for the XX-modes and the XY-modes are $626.7$ Hz and $102.6$ Hz, respectively. Overtone modes also exist at frequencies higher than these ones, and one can see that the platform thermal noise becomes significant around those resonant peak frequencies. Since the lowest resonant frequency is much above the SOGRO bandwidth, {\it e.g.,} $(0.1 \sim 10)$ Hz, however, the platform configuration having the parameters designed will be fine.

\begin{figure}[t]
\begin{center}
\includegraphics[width=0.8\textwidth]{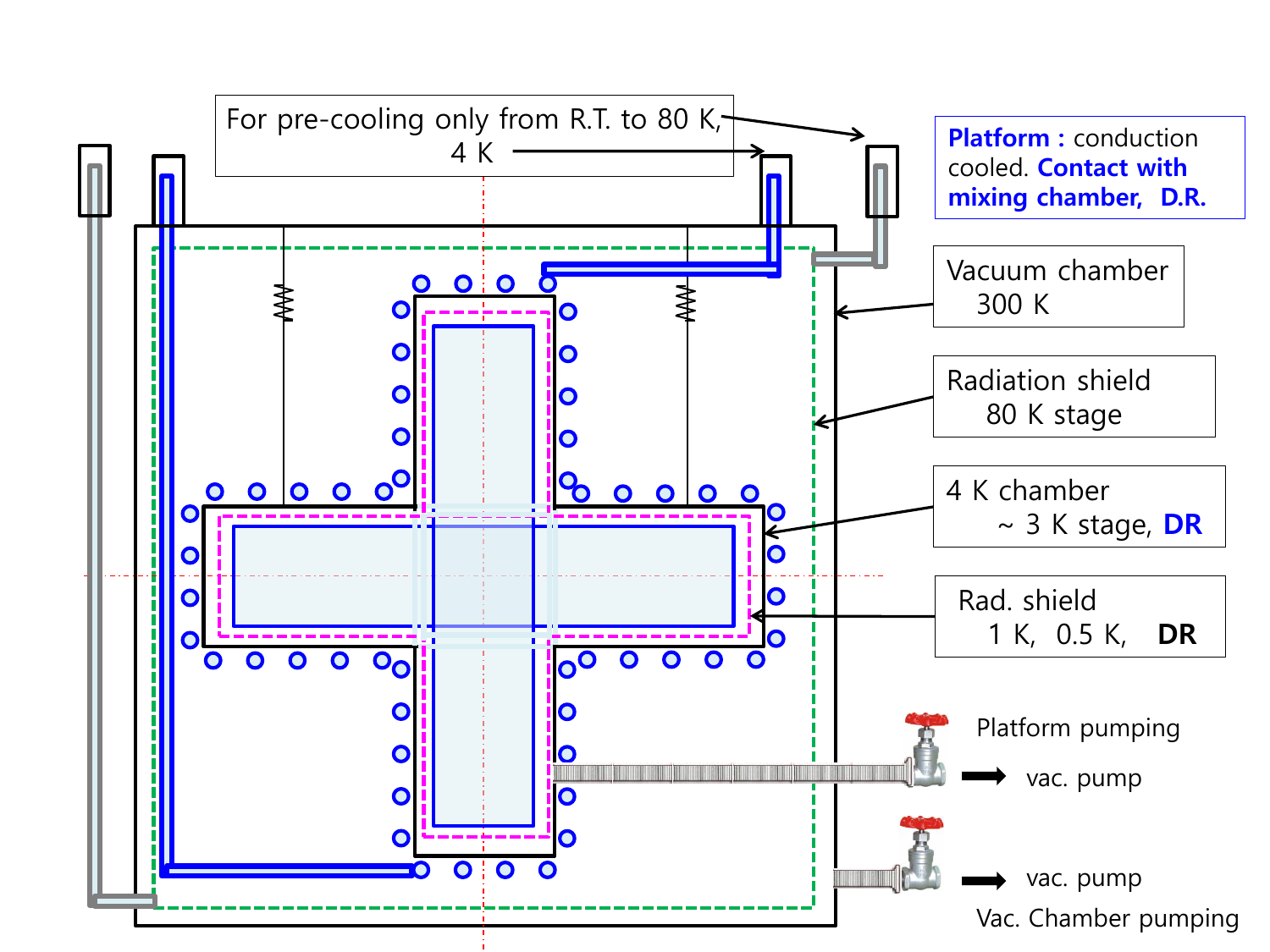}
\end{center}
\caption{Sketch of the pSOGRO's cooling system. All system is in a vacuum isolated in the cryostat. Radiation heat load from room temperature (R.T., 300 K) is shielded by 80 K liquid nitrogen shield plate. The platform is in the 4 K chamber and cooled by 4 K surface through helium heat exchange gas.}
\label{fig:cooling}
\end{figure}

\subsection{Cooling system}
\label{sec:cooling}

One of the key ideas of SOGRO is using SQUID sensors which should be operated at superconducting states. The test masses and sensing coils should also be in superconducting states. These require that SOGRO should be operated at an extremely low temperature. Thus, it is very important to test cooling methods and extract potential problems and improvements in a prototype experiment. Fig.~\ref{fig:cooling} shows the sketch of a cooling system proposed for the pSOGRO. It assumes about two tons of cold mass in total, including $600$ kg of test masses and $\sim 1$ ton of platform mass. The room temperature vacuum chamber has a height of $3.5$ m with a diameter of $3.5$ m. A $80$ K radiation shield having $3.2$ m height and $3.2$ m diameter is located inside this chamber. A $4$ K chamber having arm length of $2.8$ m and $0.7$ m diameter tube encloses three orthogonal pSOGRO platform arms. A $1$ K additional radiation shield is located inside the $4$ K chamber.

The initial cooling down from room temperature to $80$ K can be done by liquid nitrogen in the $80$ K radiation shield plate. For cooling down the platform from $80$ K to $4.2$ K, we propose a gas cooling, instead of directly immersing it in a liquid helium cryogen, in which a heat exchange gas flows in the $4$ K chamber which is in contact with liquid helium cooling tubes on the wall. This indirect cooling method avoids the noise due to the boiling of liquid helium on detector surfaces. The last step of obtaining $0.1$ K is a conduction cooling in which a dry type ${\rm He}^3$-${\rm He}^4$ dilution refrigerator ($\ge 1000$ $\mu$W at $0.1$ K) will be used.


\section{Discussion}
\label{discussion}

We have investigated the characteristics of the SOGRO detector concepts that can observe GWs in $0.1-10$ Hz and the plausible science cases with those. We consider three different concepts labeled as pSOGRO, SOGRO, and aSOGRO depending on the size and their specification.
Based on the detector characteristics, we obtain response functions for all detection channels of a SOGRO detector, given GW polarizations including even vector, scalar, breeding and longitudinal modes which may appear in alternative gravity theories. By combining responses from all channels, we conclude that SOGRO is essentially an omnidirectional GW detector. Furthermore, the tensorial nature of the SOGRO data makes it possible to localize the source position from observation with a single detector. 

SOGRO and aSOGRO with the expected sensitivities can detect IMBHs with the total mass range of about $(10^2-10^5)$ M$_{\odot}$. aSOGRO can reach up to $\sim6$ Gpc for IMBHs with ($2.5\times10^{4}-2.5\times10^{4}$) M$_{\odot}$ BBHs. This implies a detection rate of $1.0-2.5$ IMBHBs per year in the most optimistic case. For the stellar-mass black holes, the GW signals can remain in the SOGRO's frequency band for days to months, which will help to estimate parameters more accurately. Additionally, it can be a precursor to LVK observations due to its lower frequency band. However, the sensitivity of the SOGRO still needs to be improved. For the stochastic GW backgrounds, the SOGRO detector network would allow us to constrain different models of the SGWB such as the cosmological indirect limits if multiple detectors based on the SOGRO or aSOGRO design are to be realized.

In spite of its limited sensitivity, the 2 m pSOGRO can play a milestone in order to develop much larger versions such as SOGRO or aSOGRO. The detector response function obtained from the pSOGRO experiment will be one of the most important and useful information for developing SOGRO and/or aSOGRO. The design parameters for pSOGRO's platform, test masses, and SQUID sensors have been suggested, and its platform structure has been analyzed with the ANSYS software to obtain the noises. A schematic picture of the cooling system has also been suggested.

There are many technical challenges in the development of SOGRO detectors. They include Newtonian noise mitigation, highly-sensitive SQUID sensor development, civil engineering of large platforms underground, analyzing the bending effect of long arms, etc. It was shown in Ref.~\cite{PhysRevD.92.022001} that Newtonian noises originated from seismic and atmospheric density fluctuations can greatly be mitigated by optimally combining SOGRO channels with a modest number of seismometers and microphones. However, the subtraction of infra-sound noise still does not reach the required sensitivity limit. In addition, the finite size effect of the SOGRO detector might be taken into account for more accurate modeling as mentioned in Ref.~\cite{PhysRevD.92.022001}. A dedicated work to tackle on the issue of Newtonian noise mitigation is strongly called for. In particular, the Newtonian noise mitigation requires a deep underground construction at the scale larger than $\left( 50 \times 50 \times 50 \right) \, {\rm m}^3$. 

Highly sensitive SQUID sensors are another key technology required for the SOGRO GW observation. The required noise levels of the two-stage dc SQUID sensor are $20 \hbar$ and $5 \hbar$ for SOGRO and aSOGRO at $0.1$ K, respectively, as can be seen in Table~\ref{tab:DP}. It was demonstrated in Ref.~\cite{Falferi200810SQ} that a noise level of $10 \hbar$ at $\leq 0.3$ K could be achieved for a 2-stage dc SQUID at the pump frequency of $8.9$ kHz in which a commercial SQUID sensor was used~\cite{Paik2020IJMPD}. A Better noise level of $5 \hbar$ for the aSOGRO should be developed in the prototype experiment of pSOGRO. SOGRO requires a sensor circuit that is ${\cal O}(10)$ m as test masses are separated at arm length distances. The performance and quality factor of this large circuit should be examined by performing experiments when pSOGRO is realized. 

The SOGRO or aSOGRO platform including six test masses weighs around $250$ tons. According to the proposed design, the weight should be suspended by a single string or multiple strings at the center to form a pendulum as a whole. This requires that any vibrational noise produce only common motions of the ideal rigid platform through the suspension. An optimal platform designs satisfying various requirements has been suggested in Ref.~\cite{Paik2020IJMPD}, however, we note that practical civil engineering issues about how to construct and suspend the platform should further be investigated.


\section*{Acknowledgment}

This research was partially supported by the National Research Council of Science \& Technology(NST) grant by the Korean government (MSIT) (No. PCS-17-00-0000).
C.K. was supported by the Ewha Womans University Research Grant of 2023. C.P. and Y.B.B. were supported in part by IBS under Project Code No.\ IBS-R018-D1. C.P. would also like to acknowledge APCTP for its supports through activity program. Y.B.B. and E.J.S. were supported by the National Research Foundation of Korea (NRF) grants No. NRF-2021R1F1A1051269 and No. 2021R1A2C1093059 funded by the Korean government(MSIT), respectively. C.P., J.J.O. and G.K. were supported in part by the Basic Science Research Program through the National Research Foundation of Korea (NRF) funded by the Ministry of Education with grant numbers NRF-2018R1D1A1B07041004, NRF-2020R1I1A2054376 and 2022R1I1A207366012, respectively. G. K. is supported in part by National Research Foundation of Korea (NRF-2021R1A2C201247313). 
H. J. P. and R. S. N. acknowledge the support from the National Science Foundation through grant numbers PHY-1912627 and PHY-2207757.
Authors would like to acknowledge Korea Astronomy and Space Science Institute (KASI), Korea Institute of Science and Technology Information (KISTI) and National Institute for Mathematical Sciences (NIMS) for their various supports when this work was initiated.

\bibliographystyle{ptephy}
\bibliography{references}

\begin{thebibliography}{10}

\bibitem{Abbott:GW150914}
B.~P. Abbott et~al., Phys. Rev. Lett., {\bf 116}(6), 061102 (2016),
  {{arXiv:1602.03837}}.

\bibitem{aLIGO:2015}
{LIGO Scientific Collaboration} et~al., Classical and Quantum Gravity, {\bf
  32}(7), 074001 (April 2015),  {{arXiv:1411.4547}}.

\bibitem{aVirgo:2015}
F.~{Acernese} et~al., Classical and Quantum Gravity, {\bf 32}(2), 024001
  (January 2015),  {{arXiv:1408.3978}}.

\bibitem{KAGRA:2019}
{Kagra Collaboration}, T.~{Akutsu}, et~al., Nature Astronomy, {\bf 3}, 35--40
  (January 2019),  {{arXiv:1811.08079}}.

\bibitem{EPTA3:2023}
J.~{Antoniadis} et~al., arXiv e-prints, page arXiv:2306.16214 (June 2023),
  {{arXiv:2306.16214}}.

\bibitem{EPTA4:2023}
J.~{Antoniadis} et~al., arXiv e-prints, page arXiv:2306.16226 (June 2023),
  {{arXiv:2306.16226}}.

\bibitem{PPTA:2023}
Daniel~J. {Reardon} et~al., \apjl, {\bf 951}(1), L6 (July 2023),
  {{arXiv:2306.16215}}.

\bibitem{NANOGrav:2023}
Gabriella {Agazie} et~al., \apjl, {\bf 951}(1), L8 (July 2023),
  {{arXiv:2306.16213}}.

\bibitem{CPTA:2023}
Heng {Xu} et~al., Research in Astronomy and Astrophysics, {\bf 23}(7), 075024
  (July 2023),  {{arXiv:2306.16216}}.

\bibitem{LISA:2017}
Pau {Amaro-Seoane} et~al., arXiv e-prints, page arXiv:1702.00786 (February
  2017),  {{arXiv:1702.00786}}.

\bibitem{ET:2010}
M.~{Punturo} et~al., Classical and Quantum Gravity, {\bf 27}(19), 194002
  (October 2010).

\bibitem{PhysRevD.88.122003}
Jan Harms, Bram J.~J. Slagmolen, Rana~X. Adhikari, M.~Coleman Miller, Matthew
  Evans, Yanbei Chen, Holger M\"uller, and Masaki Ando, Phys. Rev. D, {\bf 88},
  122003 (Dec 2013).

\bibitem{Paik:2016aia}
Ho~Jung Paik, Cornelius~E. Griggs, M.~Vol Moody, Krishna Venkateswara,
  Hyung~Mok Lee, Alex~B. Nielsen, Ettore Majorana, and Jan Harms, Class. Quant.
  Grav., {\bf 33}(7), 075003 (2016).

\bibitem{Paik2020IJMPD}
Ho~Jung {Paik}, M.~{Vol Moody}, and Ronald~S. {Norton}, International Journal
  of Modern Physics D, {\bf 29}(4), 1940001--265 (January 2020).

\bibitem{DECIGO:2021}
Seiji {Kawamura} et~al., Progress of Theoretical and Experimental Physics, {\bf
  2021}(5), 05A105 (May 2021),  {{arXiv:2006.13545}}.

\bibitem{TOBA:2010}
Masaki {Ando}, Koji {Ishidoshiro}, Kazuhiro {Yamamoto}, Kent {Yagi}, Wataru
  {Kokuyama}, Kimio {Tsubono}, and Akiteru {Takamori}, \prl, {\bf 105}(16),
  161101 (October 2010).

\bibitem{MIGA:2018}
B.~{Canuel} et~al., Scientific Reports, {\bf 8}, 14064 (September 2018),
  {{arXiv:1703.02490}}.

\bibitem{BBO:2005}
Jeff {Crowder} and Neil~J. {Cornish}, \prd, {\bf 72}(8), 083005 (October 2005),
   {{arXiv:gr-qc/0506015}}.

\bibitem{PhysRevD.92.022001}
Jan Harms and Ho~Jung Paik, Phys. Rev. D, {\bf 92}, 022001 (Jul 2015).

\bibitem{wagoner:1979}
R.~V. Wagoner, C.~M. Will, and H.~J. Paik, Phys. Rev. D, {\bf 19}, 2325 (1979).

\bibitem{Paik:1993a}
Ho~Jung {Paik},
\newblock {Sensitivity and bandwidth of resonant-mass gravitational wave
  detectors},
\newblock In S-W {Kim}, editor, {\em Int. Workshop on Gravitation and Fifth
  Force}, pages 1--20 (1993).

\bibitem{Paik:1980qz}
H.~J. Paik, Nuovo Cim. B, {\bf 55}, 15--36 (1980).

\bibitem{Paik:1986mr}
H.~J. Paik, Phys. Rev. D, {\bf 33}, 309--318 (1986).

\bibitem{Chan:1987fv}
H.~A. Chan and H.~J. Paik, Phys. Rev. D, {\bf 35}, 3551--3571 (1987).

\bibitem{Chan:1987fw}
H.~A. Chan, M.~V. Moody, and H.~J. Paik, Phys. Rev. D, {\bf 35}, 3572--3597
  (1987).

\bibitem{2002RScI...73.3957M}
M.~Vol {Moody}, Ho~Jung {Paik}, and Edgar~R. {Canavan}, Review of Scientific
  Instruments, {\bf 73}(11), 3957--3974 (November 2002).

\bibitem{2018EPJ}
Ho~Jung Paik, EPJ Web Conf., {\bf 168}, 01005 (January 2018).

\bibitem{Nishizawa:2009bf}
Atsushi Nishizawa, Atsushi Taruya, Kazuhiro Hayama, Seiji Kawamura, and
  Masa-aki Sakagami, Phys. Rev., {\bf D79}, 082002 (2009),
  {{arXiv:0903.0528}}.

\bibitem{Cinquegrana:1993zg}
C.~Cinquegrana, P.~Rapagnani, F.~Ricci, and E.~Majorana, Phys. Rev. D, {\bf
  48}, 448--465 (1993).

\bibitem{Christensen:1992wi}
N.~Christensen, Phys. Rev. D, {\bf 46}, 5250--5266 (1992).

\bibitem{Sathyaprakash:2009xs}
B.~S. Sathyaprakash and B.~F. Schutz, Living Rev. Rel., {\bf 12}, 2 (2009),
  {{arXiv:0903.0338}}.

\bibitem{Varma:2014jxa}
Vijay Varma, Parameswaran Ajith, Sascha Husa, Juan~Calderon Bustillo, Mark
  Hannam, and Michael P\"urrer, Phys. Rev. D, {\bf 90}(12), 124004 (2014),
  {{arXiv:1409.2349}}.

\bibitem{spherical}
W.~W. Johnson and S.~M. Merkowitz, Phys. Rev. Lett., {\bf 70}, 2367--2370
  (1993).

\bibitem{minigrailweb}
{Mini-GRAIL},
\newblock \url{http://www.minigrail.nl/} (),
\newblock Accessed: 2018-05-29.

\bibitem{Hyun:2018pgn}
Young-Hwan Hyun, Yoonbai Kim, and Seokcheon Lee, Phys. Rev. D, {\bf 99}(12),
  124002 (2019),  {{arXiv:1810.09316}}.

\bibitem{Finn:2000hj}
Lee~Samuel Finn, Phys. Rev. D, {\bf 63}, 102001 (2001),  {{gr-qc/0010033}}.

\bibitem{Abbott:2020qfu}
B.~P. Abbott et~al., Living Rev. Rel., {\bf 23}(1), 3 (2020).

\bibitem{PhysRevD.51.2517}
Carl~Z. Zhou and Peter~F. Michelson, Phys. Rev. D, {\bf 51}, 2517--2545 (Mar
  1995).

\bibitem{Allen:1997ad}
Bruce Allen and Joseph~D. Romano, Phys. Rev. D, {\bf 59}, 102001 (1999),
  {{gr-qc/9710117}}.

\bibitem{GWTC1}
B.~P. {Abbott}, R.~{Abbott}, T.~D. {Abbott}, and et~al., Physical Review X,
  {\bf 9}(3), 031040 (July 2019),  {{arXiv:1811.12907}}.

\bibitem{GWTC2}
R.~{Abbott}, T.~D. {Abbott}, S.~{Abraham}, et~al., Physical Review X, {\bf
  11}(2), 021053 (April 2021),  {{arXiv:2010.14527}}.

\bibitem{GWTC2-1}
{The LIGO Scientific Collaboration} and {the Virgo Collaboration}, arXiv
  e-prints, page arXiv:2108.01045 (August 2021),  {{arXiv:2108.01045}}.

\bibitem{GWTC3}
{The LIGO Scientific Collaboration}, {the Virgo Collaboration}, and {the KAGRA
  Collaboration}, arXiv e-prints, page arXiv:2111.03606 (November 2021),
  {{arXiv:2111.03606}}.

\bibitem{Abadie:2010}
J.~Abadie et~al., Class. Quant. Grav., {\bf 27}, 173001 (2010),
  {{arXiv:1003.2480}}.

\bibitem{Merritt:2000}
David {Merritt},
\newblock {Black Holes and Galaxy Evolution},
\newblock In Francoise {Combes}, Gary~A. {Mamon}, and Vassilis {Charmandaris},
  editors, {\em Dynamics of Galaxies: from the Early Universe to the Present},
  volume 197 of {\em Astronomical Society of the Pacific Conference Series},
  page 221 (January 2000),  {{arXiv:astro-ph/9910546}}.

\bibitem{Ferrarese:2000}
Laura {Ferrarese} and David {Merritt}, \apjl, {\bf 539}(1), L9--L12 (August
  2000),  {{arXiv:astro-ph/0006053}}.

\bibitem{Gebhardt:2000}
Karl {Gebhardt}, Ralf {Bender}, Gary {Bower}, Alan {Dressler}, S.~M. {Faber},
  Alexei~V. {Filippenko}, Richard {Green}, Carl {Grillmair}, Luis~C. {Ho}, John
  {Kormendy}, Tod~R. {Lauer}, John {Magorrian}, Jason {Pinkney}, Douglas
  {Richstone}, and Scott {Tremaine}, \apjl, {\bf 539}(1), L13--L16 (August
  2000),  {{arXiv:astro-ph/0006289}}.

\bibitem{McConnell:2011}
Nicholas~J. {McConnell}, Chung-Pei {Ma}, Karl {Gebhardt}, Shelley~A. {Wright},
  Jeremy~D. {Murphy}, Tod~R. {Lauer}, James~R. {Graham}, and Douglas~O.
  {Richstone}, \nat, {\bf 480}(7376), 215--218 (December 2011),
  {{arXiv:1112.1078}}.

\bibitem{Kormendy:2013}
John {Kormendy} and Luis~C. {Ho}, Annu. Rev. Astron. Astrophys., {\bf 51}(1),
  511--653 (August 2013),  {{arXiv:1304.7762}}.

\bibitem{Patruno:2006}
A.~{Patruno}, S.~{Portegies Zwart}, J.~{Dewi}, and C.~{Hopman}, \mnras, {\bf
  370}(1), L6--L9 (July 2006),  {{arXiv:astro-ph/0602230}}.

\bibitem{Maccarone:2007}
Thomas~J. {Maccarone}, Arunav {Kundu}, Stephen~E. {Zepf}, and Katherine~L.
  {Rhode}, \nat, {\bf 445}(7124), 183--185 (January 2007),
  {{arXiv:astro-ph/0701310}}.

\bibitem{Chilingarian:2018}
Igor~V. {Chilingarian}, Ivan~Yu. {Katkov}, Ivan~Yu. {Zolotukhin}, Kirill~A.
  {Grishin}, Yuri {Beletsky}, Konstantina {Boutsia}, and David~J. {Osip}, \apj,
  {\bf 863}(1), 1 (August 2018),  {{arXiv:1805.01467}}.

\bibitem{Takekawa:2019}
Shunya {Takekawa}, Tomoharu {Oka}, Yuhei {Iwata}, Shiho {Tsujimoto}, and Mariko
  {Nomura}, \apjl, {\bf 871}(1), L1 (January 2019),  {{arXiv:1812.10733}}.

\bibitem{Woo:2019}
Jong-Hak {Woo}, Hojin {Cho}, Elena {Gallo}, Edmund {Hodges-Kluck}, Huynh Anh~N.
  {Le}, Jaejin {Shin}, Donghoon {Son}, and John~C. {Horst}, Nature Astronomy,
  {\bf 3}, 755--759 (June 2019),  {{arXiv:1905.00145}}.

\bibitem{greene:2020}
Jenny~E. {Greene}, Jay {Strader}, and Luis~C. {Ho}, Annu. Rev. Astron.
  Astrophys., {\bf 58}, 257--312 (August 2020),  {{arXiv:1911.09678}}.

\bibitem{Ajith:2011}
P.~Ajith et~al., Phys. Rev. Lett., {\bf 106}, 241101 (2011),
  {{arXiv:0909.2867}}.

\bibitem{Planck:2016}
P.~A.~R. Ade et~al., Astron. Astrophys., {\bf 594}, A13 (2016),
  {{arXiv:1502.01589}}.

\bibitem{Giersz:2016}
M.~{Giersz}, N.~{Leigh}, A.~{Hypki}, A.~{Askar}, and N.~{L{\"u}tzgendorf}, Mem.
  Soc. Astron. Italiana, {\bf 87}, 555 (January 2016),  {{arXiv:1607.08384}}.

\bibitem{Fregeau:2006}
John~M. {Fregeau}, Shane~L. {Larson}, M.~Coleman {Miller}, Richard
  {O'Shaughnessy}, and Frederic~A. {Rasio}, \apjl, {\bf 646}(2), L135--L138
  (August 2006),  {{arXiv:astro-ph/0605732}}.

\bibitem{Amaro-Seoane:2010}
Pau {Amaro-Seoane} and Luc{\'\i}a {Santamar{\'\i}a}, \apj, {\bf 722}(2),
  1197--1206 (October 2010),  {{arXiv:0910.0254}}.

\bibitem{Portegies_Zwart:1999}
S.~F. {Portegies Zwart}, J.~{Makino}, S.~L.~W. {McMillan}, and P.~{Hut}, \aap,
  {\bf 348}, 117--126 (August 1999),  {{arXiv:astro-ph/9812006}}.

\bibitem{Ebisuzaki:2001}
Toshikazu {Ebisuzaki}, Junichiro {Makino}, Takeshi~Go {Tsuru}, Yoko {Funato},
  Simon {Portegies Zwart}, Piet {Hut}, Steve {McMillan}, Satoki {Matsushita},
  Hironori {Matsumoto}, and Ryohei {Kawabe}, \apjl, {\bf 562}(1), L19--L22
  (November 2001),  {{arXiv:astro-ph/0106252}}.

\bibitem{Portegies_Zwart:2002}
Simon~F. {Portegies Zwart} and Stephen L.~W. {McMillan}, \apj, {\bf 576}(2),
  899--907 (September 2002),  {{arXiv:astro-ph/0201055}}.

\bibitem{Portegies_Zwart:2004}
Simon~F. {Portegies Zwart}, Holger {Baumgardt}, Piet {Hut}, Junichiro {Makino},
  and Stephen L.~W. {McMillan}, \nat, {\bf 428}(6984), 724--726 (April 2004),
  {{arXiv:astro-ph/0402622}}.

\bibitem{Gurkan:2004}
M.~Atakan {G{\"u}rkan}, Marc {Freitag}, and Frederic~A. {Rasio}, \apj, {\bf
  604}(2), 632--652 (April 2004),  {{arXiv:astro-ph/0308449}}.

\bibitem{Madau:2000}
Piero {Madau} and Lucia {Pozzetti}, \mnras, {\bf 312}(2), L9--L15 (February
  2000),  {{arXiv:astro-ph/9907315}}.

\bibitem{Steidel:1999}
Charles~C. {Steidel}, Kurt~L. {Adelberger}, Mauro {Giavalisco}, Mark
  {Dickinson}, and Max {Pettini}, \apj, {\bf 519}(1), 1--17 (July 1999),
  {{arXiv:astro-ph/9811399}}.

\bibitem{Blain:1999}
A.~W. {Blain}, J.~P. {Kneib}, R.~J. {Ivison}, and Ian {Smail}, \apjl, {\bf
  512}(2), L87--L90 (February 1999),  {{arXiv:astro-ph/9812412}}.

\bibitem{Porciani:2001}
Cristiano {Porciani} and Piero {Madau}, \apj, {\bf 548}(2), 522--531 (February
  2001),  {{arXiv:astro-ph/0008294}}.

\bibitem{McLaughlin:1999}
Dean~E. {McLaughlin}, Astron. J., {\bf 117}(5), 2398--2427 (May 1999),
  {{arXiv:astro-ph/9901283}}.

\bibitem{Freitag:2006a}
Marc {Freitag}, Frederic~A. {Rasio}, and Holger {Baumgardt}, \mnras, {\bf
  368}(1), 121--140 (May 2006),  {{arXiv:astro-ph/0503129}}.

\bibitem{Freitag:2006b}
Marc {Freitag}, M.~Atakan {G{\"u}rkan}, and Frederic~A. {Rasio}, \mnras, {\bf
  368}(1), 141--161 (May 2006),  {{arXiv:astro-ph/0503130}}.

\bibitem{Zhang:1999}
Qing {Zhang} and S.~Michael {Fall}, \apjl, {\bf 527}(2), L81--L84 (December
  1999),  {{arXiv:astro-ph/9911229}}.

\bibitem{Amaro_Seoane:2006}
Pau {Amaro-Seoane} and Marc {Freitag}, \apjl, {\bf 653}(1), L53--L56 (December
  2006),  {{arXiv:astro-ph/0610478}}.

\bibitem{sedda:2020}
Manuel {Arca Sedda} et~al., Classical and Quantum Gravity, {\bf 37}(21), 215011
  (November 2020),  {{arXiv:1908.11375}}.

\bibitem{gerosa:2017}
Davide {Gerosa} and Emanuele {Berti}, \prd, {\bf 95}(12), 124046 (June 2017),
  {{arXiv:1703.06223}}.

\bibitem{Shinkai:2017}
Hisa-Aki Shinkai, Nobuyuki Kanda, and Toshikazu Ebisuzaki, Astrophys. J., {\bf
  835}(2), 276 (2017),  {{arXiv:1610.09505}}.

\bibitem{nakamura:2016}
Takashi {Nakamura} et~al., Progress of Theoretical and Experimental Physics,
  {\bf 2016}(9), 093E01 (September 2016),  {{arXiv:1607.00897}}.

\bibitem{carr:2020}
Bernard {Carr} and Florian {K{\"u}hnel}, Annual Review of Nuclear and Particle
  Science, {\bf 70}, 355--394 (October 2020),  {{arXiv:2006.02838}}.

\bibitem{shandera:2018}
Sarah {Shandera}, Donghui {Jeong}, and Henry~S. {Grasshorn Gebhardt}, \prl,
  {\bf 120}(24), 241102 (June 2018),  {{arXiv:1802.08206}}.

\bibitem{Nitz:2021}
Alexander~H. {Nitz} and Yi-Fan {Wang}, \prl, {\bf 127}(15), 151101 (October
  2021),  {{arXiv:2106.08979}}.

\bibitem{thrane_sensitivity_2013}
Eric Thrane and Joseph~D. Romano, Physical Review D, {\bf 88}(12), 124032
  (December 2013).

\bibitem{gwplotter}
Vuk Mandic and Erik Floden,
\newblock An interactive plotter for energy spectrum of stochastic
  gravitational wave backgrounds from various theoretical models,
\newblock \url{https://homepages.spa.umn.edu/gwplotter/} ().

\bibitem{PhysRevD.104.022004}
R.~Abbott et~al., Phys. Rev. D, {\bf 104}, 022004 (Jul 2021).

\bibitem{PhysRevX.6.011035}
Paul~D. Lasky et~al., Phys. Rev. X, {\bf 6}, 011035 (Mar 2016).

\bibitem{Cornish_2017}
Neil Cornish and Travis Robson, Journal of Physics: Conference Series, {\bf
  840}, 012024 (may 2017).

\bibitem{planck2018}
Y.~Akrami et~al., A\&A, {\bf 641}, A10 (2020).

\bibitem{PhysRevApplied.8.064024}
C.~E. Griggs, M.~V. Moody, R.~S. Norton, H.~J. Paik, and K.~Venkateswara, Phys.
  Rev. Applied, {\bf 8}, 064024 (Dec 2017).

\bibitem{Falferi200810SQ}
P.~Falferi, M.~Bonaldi, Massimo Cerdonio, R.~Mezzena, G.~Prodi, A.~Vinante, and
  S.~Vitale, Applied Physics Letters, {\bf 93}, 172506 -- 172506 (11 2008).

\end{thebibliography}


\newpage

\appendix

\section{Response functions for extra polarizations of gravitational waves}
\label{appendixA}

Alternative theories of gravity could allow extra polarization degrees of freedom for gravitational waves in addition to the plus and cross ones in general relativity. Here, the response functions for such extra modes are summarized for the laser interferometer and SOGRO detectors. First of all, the three unit vectors mentioned in Sect.~\ref{sec:response} are explicitly given by
\begin{align}
\hat{u} &= ( \sin \phi \cos \psi + \cos \theta \cos \phi \sin \psi, -\cos \phi \cos \psi + \cos \theta \sin \phi \sin \psi, -\sin \theta \sin \psi ), \nonumber \\
\hat{v} &= ( \cos \theta \cos \phi \cos \psi - \sin \phi \sin \psi, \cos \theta \sin \phi \cos \psi + \cos \phi \sin \psi, - \sin \theta \cos \psi ),  \nonumber \\
\hat{n} &= ( \sin \theta \cos \phi, \sin \theta \sin \phi, \cos \theta ),
\label{unit-vectors}
\end{align}
where $\theta$ and $\phi$ are the polar and azimuthal angles of the propagating direction $\hat{n}$, respectively, and $\psi$ are the polarization angle.

For the laser interferometer, the response functions for GWs having vector, breathing and longitudinal polarizations are given by
\begin{eqnarray}
    F_{x}(\theta,\phi,\psi) &=& \sin \theta \sin 2\phi \cos \psi + \frac{1}{2}\sin 2\theta \cos 2\phi \sin \psi, \nonumber \\
    F_{y}(\theta,\phi,\psi) &=& \frac{1}{2}\sin 2\theta \cos 2\phi \cos \psi - \sin \theta \sin 2\phi \sin \psi ,  \nonumber \\
    F_{b}(\theta,\phi,\psi) &=& -\frac{1}{2}\sin^2 \theta \cos 2\phi ,  \nonumber  \\
    F_{l}(\theta,\phi,\psi) &=& \frac{1}{\sqrt{2}} \sin^2 \theta \cos 2\phi ,
\label{interferometer-vbl-modes}
\end{eqnarray}
respectively.
For SOGRO channels, on the other hand, they become
\begin{eqnarray}
    F_{x}^{(11)} = \sin \theta \sin 2\phi \cos \psi + \sin 2\theta \cos^2 \phi \sin \psi ,  \nonumber \\
    F_{x}^{(22)} = -\sin \theta \sin 2\phi \cos \psi + \sin 2\theta \sin^2 \phi \sin \psi ,  \nonumber \\
    F_{x}^{(12)} = -\sin \theta \cos 2\phi \cos \psi + \frac{1}{2} \sin 2\theta \sin 2\phi \sin \psi ,  \nonumber \\
    F_{y}^{(11)} = -\sin \theta \sin 2\phi \sin \psi + \sin 2\theta \cos^2 \phi \cos \psi ,  \nonumber \\
    F_{y}^{(22)} = \sin 2\theta \sin^2 \phi \cos \psi + \sin \theta \sin 2\phi \sin \psi ,  \nonumber \\
    F_{y}^{(12)} = \frac{1}{2} \sin 2\theta \sin 2\phi \cos \psi + \sin \theta \cos 2\phi \sin \psi ,  \nonumber \\
    F_{b}^{(11)} = \sin^2 \phi +\cos^2 \theta \cos^2 \phi , \quad
    F_{b}^{(12)} = -\frac{1}{2} \sin^2 \theta \sin 2\phi ,  \nonumber  \\
    F_{l}^{(11)} = \sqrt{2} \sin^2 \theta \cos^2 \phi , \quad
    F_{l}^{(12)} = \frac{1}{\sqrt{2}} \sin^2 \theta \sin 2\phi ,
\label{SOGRO-vbl-modes}
\end{eqnarray}
respectively.
Response functions of a SOGRO detector were defined in Eq.~\eqref{SOGROoutput}.

\end{document}